\documentclass[12pt,preprint]{aastex}
\newcommand{\beq}{\begin{equation}}
\newcommand{\eeq}{\end{equation}}
\def\stacksymbols #1#2#3#4{\def\theguybelow{#2}
        \def\verticalposition{\lower#3pt}
        \def\spacingwithinsymbol{\baselineskip0pt\lineskip#4pt}
        \mathrel{\mathpalette\intermediary#1}}
\def\intermediary #1#2{\verticalposition\vbox{\spacingwithinsymbol
        \everycr={}\tabskip0pt
        \halign{$\mathsurround0pt#1\hfil##\hfil$\crcr#2\crcr
                \theguybelow\crcr}}}

\catcode`\@=11
\def\gsim{\ifmmode{\mathrel{\mathpalette\@versim>}}
    \else{$\mathrel{\mathpalette\@versim>}$}\fi}
\def\lsim{\ifmmode{\mathrel{\mathpalette\@versim<}}
    \else{$\mathrel{\mathpalette\@versim<}$}\fi}
\def\@versim#1#2{\lower 2.9truept \vbox{\baselineskip 0pt \lineskip 
    0.5truept \ialign{$\m@th#1\hfil##\hfil$\crcr#2\crcr\sim\crcr}}}
\catcode`\@=12

\def\brem{bremsstrahlung$\;\,$}

\def\Lsun{L_{\odot}}
\def\Msun{M_{\odot}}

\def\eps{\epsilon}
\def\epsz{\epsilon_0}
\def\epsII{\eps_{\rm II}}
\def\epsw{\eps_{\rm w}}
\def\epswM{\epsw^{\rm M}}
\def\epsj{\eps_{\rm j}}

\def\epsA{\eps_{\rm EM}}
\def\tauII{\tau_{\rm II}}
\def\tauopt{\tau_{\rm opt}}

\def\taul{\tau_{\rm *l}}
\def\tauh{\tau_{\rm *h}}
\def\tlagd{\tau_{\rm d}}
\def\tlagi{\tau_{\rm i}}

\def\lb{L_{\rm B}}
\def\lbh{L_{\rm BH}}

\def\ledd{L_{\rm Edd}}
\def\ldwin{L_{\rm dw}}
\def\lj{L_{\rm j}}

\def\lbhefopt{L_{\rm BH,opt}^{\rm eff}}

\def\lduv{L_{\rm d,UV}}
\def\ldopt{L_{\rm d,opt}}

\def\Pwj{P_{\rm wj}}
\def\Ywj{Y_{\rm wj}}
\def\mast{M_*}

\def\mgas{M_{\rm gas}}
\def\Min{M_{\rm inf}}
\def\MII{M_{\rm II}}

\def\mbh{M_{\rm BH}}

\def\Mdg{M_{\rm dg}}
\def\Mds{M_{\rm d*}}
\def\Mdsl{M_{\rm dl*}}
\def\Mdsh{M_{\rm dh*}}
\def\Mdw{M_{\rm dw}}
\def\Mrem{M_{\rm rem}}
\def\Mfid{M_{\rm fid}}
\def\Medd{M_{\rm Edd}}
\def\Mj{M_{\rm j}}
\def\Mwj{M_{\rm wj}}

\def\RM{{\cal R}}
\def\ML{\Upsilon_*}
\def\mdot{\dot\mbh}
\def\dmin{\dot M_{1}}
\def\dmineff{\dot M_1^{\rm eff}}

\def\dMfid{\dot\Mfid}
\def\rhos{\rho_*}

\def\re{R_{\rm e}}
\def\rs{r_*}

\def\Rd{R_{\rm d}}
\def\Rc{R_{\rm c}}
\def\Rwj{R_{\rm wj}}

\def\tx{T_{\rm X}}

\def\cs{c_{\rm s}}
\def\ceq{c_{\rm eq}}

\def\fh{f_{\rm h}}

\def\freml{f_{\rm rem,l}}
\def\fremh{f_{\rm rem,h}}

\def\etaD{\eta_{\rm d}}
\def\etaw{\eta_{\rm w}}
\def\etawM{\etaw^{\rm M}}
\def\etaj{\eta_{\rm j}}
\def\etas{\eta_*}

\def\vw{v_{\rm w}}
\def\vj{v_{\rm j}}
\def\vwj{v_{\rm wj}}
\def\vff{v_{\rm ff}}
\def\DOmew{\Delta\Omega_{\rm w}}
\def\DOmej{\Delta\Omega_{\rm j}}
\def\DOmec{\Delta\Omega_{\rm c}}
\def\mw{m_{\rm w}}
\def\mj{m_{\rm j}}
\def\Thc{\Theta_{\rm c}}

\def\Pism{P_{\rm ISM}}

\def\sigap{\sigma_{\circ}}
\def\sigast{\sigma_*}

\def\RBZzd{RB0$_{02}$}

\def\t15{t_{15}}

\slugcomment{Accepted version - April 20, 2009}

\lefthead{L.~Ciotti, J.P.~Ostriker, and D.~Proga}
\righthead{BH feedback on the ISM of ellipticals}

\begin{document}
\title{Feedback from central black holes in elliptical galaxies.\\ I:
       models with either radiative or mechanical feedback but not both}

\author{Luca Ciotti\altaffilmark{1}, 
Jeremiah P. Ostriker\altaffilmark{2,3} and Daniel Proga\altaffilmark{4}}
\affil{$^1$Department of Astronomy, University of Bologna,
via Ranzani 1, I-40127, Bologna, Italy} 
\affil{$^2$Princeton University Observatory, Princeton, NJ, USA}
\affil{$^3$IoA, Cambridge, UK}
\affil{$^4$Department of Physics and Astronomy, University of Nevada, Las Vegas,
           NV, USA}

\begin{abstract} 

  The importance of the radiative feedback from massive black holes at
  the centers of elliptical galaxies is not in doubt, given the well
  established relations among electromagnetic output, black hole mass
  and galaxy optical luminosity.  In addition, feedback due to
  mechanical and thermal deposition of energy from jets and winds
  emitted by the accretion disk around the central black hole is also
  expected to occur and has been included in the work of several
  investigators.  In this paper we improve and extend the accretion
  and feedback physics explored in our previous papers to include also
  a physically motivated model of mechanical feedback, in
  addition to radiative effects.  In particular, we study the
  evolution of an isolated elliptical galaxy with the aid of a
  high-resolution 1-D hydrodynamical code, where the cooling and
  heating functions include photoionization and Compton effects, and
  restricting to models which include only radiative or only
  mechanical feedback (in the form of nuclear winds).  We confirm that
  for Eddington ratios above 0.01 both the accretion and radiative
  output are forced by feedback effects to be in burst mode, so that
  strong intermittencies are expected at early times, while at low
  redshift the explored models are characterized by smooth, very
  sub-Eddington mass accretion rates punctuated by rare outbursts.
  However, the explored models always fail some observational tests.
  If we assume the high mechanical efficiency of $10^{-2.3}$ adopted
  by some investigators, we find that most of the gas is ejected from
  the galaxy, the resulting X-ray luminosity is far less than is
  typically observed and little SMBH growth occurs.  But models with
  low enough mechanical efficiency to accomodate satisfactory SMBH
  growth tend to allow too strong cooling flows and leave galaxies at
  $z=0$ with E+A spectra more frequently than is observed. In a
  surprising conclusion we find that both types of feedback are
  required. Radiative heating over the inner few kpc is needed to
  prevent calamitous cooling flows, and mechanical feedback from AGN
  winds, which affects primarily the inner few hundred pc, is needed
  to moderate the luminosity and growth of the central SMBH. Models
  with combined feedback are explored in a forthcoming paper.

\end{abstract}


\section{Introduction}

All massive bulges and elliptical galaxies contain massive black holes
at their center (hereafter SMBHs, e.g., see Kormendy \& Richstone
1995, de Zeeuw 2001, Ferrarese \& Ford 2005), and when gas is added to
the central regions for any reason the SMBH will accrete and emit
energy.  It is also clear that SMBHs have played an important role in
the processes of galaxy formation and evolution (e.g., see Silk \&
Rees 1998; Fabian 1999; Burkert \& Silk 2001; Cavaliere \& Vittorini
2002; King 2003; Wyithe \& Loeb 2003; Haiman, Ciotti \& Ostriker 2004;
Granato et al. 2004; Sazonov et al. 2005; Murray, Quataert \& Thompson
2005; Di Matteo, Springel \& Hernquist 2005; Begelman \& Nath 2005;
Hopkins et al. 2006; Croton et al. 2006; Pipino, Silk \& Matteucci
2008), as strongly supported by the remarkable correlations found
between host galaxy properties and the masses of their SMBHs (e.g.,
see Magorrian et al. 1998, Ferrarese \& Merritt 2000, Gebhardt et
al. 2000, Yu \& Tremaine 2002, McLure \& Dunlop 2002, Graham et
al. 2003, Marconi \& Hunt 2003, see also Somerville 2008, Ciotti 2009).

This basic fact leads immediately to a set of interrelated questions
which must be addressed before we can understand either the resultant
masses of the central SMBHs or the co-dependency of AGN feedback and
galactic evolution, i.e. $a)$ what physical processes control the SMBH
accretion rate?  For a given accretion rate, what form does the energy
feedback take (photons, nuclear winds and jets)?  $b)$ How do these
three forms of feedback, in turn, affect the accretion rate?

In our prior papers (Ciotti \& Ostriker 1997,2001,2007, hereafter
CO97, CO01 and CO07, and Ostriker \& Ciotti 2005, hereafter OC05), we
have focussed on the radiative, electromagnetic (EM) component, since
that is most easily observed and our knowledge of it is most certain
(e.g., see Sazonov, Ostriker \& Sunyaev 2004; Sazonov et al. 2007,
2008), and we already provided indications that accretion feedback is
the obvious solution to several related problems, such as the long
lasting ``cooling-flow'' problem, and the fate of the large amounts of
gas injected into the galaxy by the passively evolving stellar
populations (e.g., Peterson \& Fabian 2006, see also Binney 2001).
This conclusion is also supported by other investigations implementing
physically motivated feedback mechanisms (e.g., see Tabor \& Binney
1993, Binney \& Tabor 1995, Omma et al. 2004, Churazov et al. 2005,
Antonuccio Delogu \& Silk 2008), and the computed solutions are
characterized by relaxation oscillations (e.g. Cowie, Ostriker \&
Stark 1978; Milosavljevic et al.  2008).  The general emerging picture
is that energy output (radiative or mechanical) from the central SMBH
pushes matter out, the accretion rate drops precipitously and the
expanding matter drives shocks into the galactic gas.  
Then the resulting hot bubble ultimately cools radiatively (it
is thermally unstable) and the consequent infall leads to renewed
accretion; the cycle repeats, with the galaxy being seen alternately
as an AGN/starburst for a small fraction of the time and as a
``normal'' elliptical hosting an incipient cooling catastrophe for
much longer intervals. Nowadays, several observations support the
finding that accretion on central SMBHs is in fact a highly unsteady
phenomenon (e.g., see Martini 2004, Goncalves, Steidel \& Pettini
2007; Prochaska \& Hennawi 2008, Hopkins \& Hernquist 2008).

However, although our previous work has stressed the significance of
{\it radiative} heating near to the SMBH (in addition to the well
known Eddington momentum input, e.g. see Dorodnitsyn, Kallman, \&
Proga 2008; Shi \& Krolik 2008), other investigators (e.g., Binney \&
Tabor 1995, Tabor \& Binney 1993, Begelman \& Nath 2005, Begelman \&
Ruszkowski 2005, Di Matteo, Springel \& Hernquist 2005, Springel, Di
Matteo \& Hernquist 2007) have focussed on the also important but
highly uncertain {\it mechanical} feedback\footnote{Note that also a
  purely radiative feedback produces ultimately some form of
  mechanical feedback, in the form of shock waves, e.g. CO01 and CO07,
  or in the form or radiation driven winds as found in 2D and 3D
  simulations (Proga 2007, Proga et al. 2008b, Kurosawa \& Proga
  2008ab).}.  An obvious source of mechanical feedback is certainly
represented by the radiatively driven winds from the broadline regions
(BLRs), whose parameters are well observed (e.g., Chartas et al. 2003,
2007; Crenshaw, Kraemer, \& George 2003; Blustin et al. 2007; Hamann
et al. 2008), so that their energy, momentum, and mass input can be
added to the codes in a fairly direct way. Although the overall
efficiency of such inputs appears to be modest, this component of
feedback couples with great effectiveness to the ambient gas. In
addition, highly collimated jets, especially at low Eddington ratios
are observed to put out energy in amounts comparable to the EM output,
but it is not clear how efficiently such narrow jets can couple to the
ambient fluid (see Vernaleo \& Reynolds 2006).

The purpose of the current paper is to introduce a {\it physically
  based} modelling of mechanical energy input to our code, and to
supplement the detailed treatment of radiative effects, in a fashion
that is consistent with current theory and guided by observations. At
the same time, we further improve the galaxy models used for the
simulations, both in the stellar and dark matter distributions,
following the results of the latest observational works.  In
particular, in this paper we restrict our exploration to purely
radiative and purely mechanical feedback, in order to better
understand the specific properties of the two mechanisms when
considered separately.  Other aspects of purely mechanical models are
investigated in detail in Shin, Ostriker \& Ciotti (in preparation,
hereafter Paper II), while we reserve to a third paper (Ciotti \&
Ostriker, in preparation, hereafter Paper III) the discussion of
models in which both radiative and mechanical feedback effects are
included in the hydrodynamical code. Finally, the X-ray observational
properties of the most successful models will be presented in
Pellegrini, Ciotti \& Ostriker (in preparation, hereafter Paper
IV. See also Pellegrini, Ciotti \& Ostriker 2009), to be compared with
the results of observational works (e.g., O'Sullivan, Ponman, \&
Collins, 2003; Diehl \& Statler 2008).

We continue to find that the situation is inherently
complex. Mechanical energy output of the AGN is lower than the EM
output but couples more effectively to ambient gas and thus is optimal
in the innermost few hundred parsecs in shielding the SMBH from
excessive accretion. But the radiation heating input to the galaxy is
carried primarily by long mean free path (X-ray) photons and is most
effective in heating the inner several kpc and thus moderating or
forestalling cooling catastrophes in the ambient ISM of the galaxy.
We tentatively conclude the BLR winds do have an essential effect in
regulating SMBH accretion and, with the radiation heating feedback,
cooperate in removing significant amounts of gas from giant
ellipticals (see Schawinski et al. 2008) but that the jets largely
escape from isolated galaxies. However, for Brightest Cluster
Galaxies, the energetic jets can provide significant input to the
cluster gas - thereby helping to prevent or moderate cluster cooling
flow catastrophes (e.g., Voit \& Donahue 2005; Peterson \& Fabian
2006; Rafferty, McNamara \& Nulsen 2008; McCarthy et al. 2008).
 
The paper is organized as follows. In Section 2 we describe how the
new galaxy models adopted in the simulations are built, the details of
the input physics (focusing in particular on the physics of accretion
disk and on the mechanical feedback), and their numerical
implementation.  In Section 3 we present two different classes of
models, both of them presenting some interesting aspects but overall
failing to reproduce the situation observed in real galaxies. We thus
have a first class of models in which only radiative heating is taken
into account, and a second class in which only mechanical feedback is
considered. Finally, in Section 4 we discuss the main results
obtained.

\section{The models}

The galaxy models and the input physics adopted for the simulations
have been improved with respect to our previous explorations. In the
following we describe the new ingredients of the current models, while
for the unchanged parts we refer to our previous papers: a comparative
summary of the present and past treatments is given in Table 1.

\begin{deluxetable}{lccccccc}
\rotate
\tablecaption{Synoptic table of feedback model development}
\tabletypesize{\scriptsize}
\tablewidth{0pt}
\tablehead{\colhead{Input Physics}&
\colhead{CO97}&
\colhead{CO01}&
\colhead{OC05}&
\colhead{CO07}&
\colhead{PaperI [Radiative]}&
\colhead{PaperI [Mechanical]}&
\colhead{PaperIII}\\
\colhead{(1)}&
\colhead{(2)}&
\colhead{(3)}&
\colhead{(4)}& 
\colhead{(5)}&
\colhead{(6)}&
\colhead{(7)}&
\colhead{(8)}
}
\startdata
{\bf Galaxy model}    
& King+Q.I.H.
& King+Q.I.H.
& King+Q.I.H.
& Hernquist+Hernquist
& Jaffe in S.I.S. 
& Jaffe in S.I.S.  
& Jaffe in S.I.S.\\
{\bf Star formation}  
& No         
& No         
& No            
& Yes                
& Yes             
& Yes              
& Yes\\
\hline
{\bf Circumnuclear disk}        
& No         
& No         
& No           
& Yes                
& Yes   
& Yes        
& Yes\\
{\bf ADAF}     
& No        
& \tablenotemark{a}Yes    
& No            
& No                 
& Yes          
& Yes   
& Yes\\
\hline
{\bf Heating and Cooling}\\
Compton         
& \tablenotemark{b}Yes    
& \tablenotemark{b}Yes    
& Yes      
& Yes    
& Yes     
& \tablenotemark{c}Only cooling            
& Yes\\
Photoionization 
& No         
& No         
& Yes           
& Yes                
& Yes
& \tablenotemark{c}Only cooling               
& Yes\\
\hline
{\bf Radiation Pressure}\\
e$^-$ scattering 
& Yes        
& Yes        
& Yes           
& Yes                
& Yes 
& \tablenotemark{d}No    
& Yes\\
Photoionization  
& No         
& No         
& No            
& Yes   
& Yes                 
& \tablenotemark{d}No    
& Yes\\
Dust            
& No         
& No         
& No            
& Yes    
& Yes                 
& \tablenotemark{d}No    
& Yes\\
\hline
{\bf Mechanical Feedback}\\
AGN wind        
& No         
& No         
& No            
& \tablenotemark{d}No             
& \tablenotemark{d}No    
& Yes      
& Yes \\
AGN jet         
& No         
& No         
& No            
& No                 
& \tablenotemark{d}No   
& \tablenotemark{d}No    
& \tablenotemark{d}No        \\
\enddata
\tablecomments{Different names indicate the papers as in the
  text. Paper I indicates this paper; Paper III; Q.I.H.=
  Quasi-Isothermal Halo; S.I.S.= Singular Isothermal Sphere: in the
  present models the stellar density is immersed in a dark matter halo
  so that the total density profile is proportional to $r^{-2}$.}
\tablenotetext{a}{Only a few test models were computed.}
\tablenotetext{b}{In CO97 a very high Compton temperature $\tx$ for
  the emitted accretion luminosity spectral distribution was used,
  while in CO01 a large range of values for $\tx$ was explored.
  Starting from OC05 the value of $\tx\simeq 2\,10^7$ K was fixed
  according to the observational estimates of Sazonov, Ostriker \&
  Sunyaev (2004).}
\tablenotetext{c}{Heating computed but not added to the hydrodynamics.}
\tablenotetext{d}{Computed but not added to the hydrodynamics.}
\end{deluxetable}

\subsection{Structure and internal dynamics}

In CO97 and CO01 the galaxy models utilized a King (1972) stellar
distribution plus a quasi-isothermal dark matter halo, in line with
the models then used for cooling-flow studies. However, the existence
of large cores of constant surface brightness has been clearly ruled
out, as HST observations have shown how the central surface brightness
profile is described by a power-law as far in as it can be observed,
i.e. to $\sim 10$ pc from the center for Virgo ellipticals (e.g., see
Jaffe et al. 1994, Faber et al. 1997, Lauer et al. 2005). In
Pellegrini \& Ciotti (1998) and in CO07 a stellar density distribution
described by the more appropriate Hernquist (1990) model has been
adopted, and also the dark matter halo was described by an Hernquist
profile, which is quite similar in its central regions to the dark
matter halos obtained from cosmological simulations (e.g., Dubinski \&
Carlberg 1991; Navarro, Frenk \& White 1997; Fukushige \& Makino
1997).  

However, the Hernquist profile is characterized (as the de Vaucouleurs
$R^{1/4}$ density profile) by a sizable central depression in the
isotropic velocity dispersion, which is not observed.  This problem is
now fixed. In fact, the velocity dispersion of the Jaffe (1983)
profile is monotonically decreasing. In addition, observations support
the idea that ellipticals are characterized by a {\it total} density
distribution well described by a $r^{-2}$ profile (e.g., see Treu \&
Koopmans 2002,2004; Rusin et al. 2003, Rusin \& Kochanek 2005,
Koopmans et al. 2006, Gavazzi et al. 2007, Czoske et al. 2008; Dye et
al. 2008).  For these reasons, here we adopt Jaffe stellar models
embedded in a dark halo so that the total mass profile decreases as
$r^{-2}$. Therefore, the stellar density profile of the galaxy models
is
\beq
\rhos  = {\mast\rs\over 4\pi r^2(\rs+r)^2},
\eeq
where $\mast$ and $\rs$ are the total stellar mass and the
scale-length of the galaxy, respectively; we recall that for the Jaffe
model the effective radius is $\re=\rs/0.7447$. The total density
profile is then given by
\beq
\rho_{\rm T}  = {\RM \mast\over 4\pi \rs r^2},
\eeq
where the dimensionless factor $\RM\geq 1$ controls the amount of dark
matter contained within the half-mass radius of the stellar component.
With this choice, the circular velocity of the model is given by
$v_{\rm c}=\sqrt{G\RM\mast/\rs}$.  The presence of significant amounts
of dark matter in the central regions of normal galaxies remains
controversial (e.g., see Binney \& Evans 2001), but in general
observational studies in the optical (e.g., Saglia et al. 1993; Bertin
et al. 1994; Cappellari et al. 2006, Douglas et al. 2007) and in
X-rays (e.g. Fabian et al. 1986, Humphrey et al. 2006) seems to
indicate that dark matter begins to be dynamically important at
$2-3\re$.  For these reasons we fix $\RM$ to the minimum admissible
value $\RM=1$, corresponding to the same amount of dark and visible
matter within the spatial half-mass radius of the stellar
distribution.

All the dynamical and phase-space properties of the resulting
two-component galaxy models are given in Ciotti, Morganti \& de Zeeuw
(2008), and here we report only the formulae for the quantities of
interest: in particular, the model central projected velocity
dispersion $\sigap$ (obtained by solving and projecting the Jeans
equations under the assumption of orbital isotropy) is given by
\beq 
\sigap ={v_{\rm c}\over \sqrt{2}}.
\eeq 
Note that we do not consider the effect of $\mbh$ on $\sigap$, in
accordance with estimated values for the radius of the SMBH sphere of
influence (e.g., Riciputi et al. 2005).  The parameters describing the
galaxy model are determined following CO07, i.e., first we assign a
value for the central velocity dispersion $\sigap$, and we determine
the galaxy {\it present day} blue luminosity $\lb$ and effective
radius $\re$ from the Faber-Jackson and the Fundamental Plane relations
(eqs.~[3]-[4] in CO07).  We then fix the relative amount of dark
matter to stars within $\re$ by assigning $\RM$, thus determining the
total stellar mass $\mast$ of the galaxy and finally the stellar
mass-to-light ratio $\ML\equiv \mast/\lb$.

An important ingredient in the energetics of the gas flows, namely the
thermalization of the stellar mass losses due to the stellar velocity
dispersion (e.g., see Parriott \& Bregman 2008), 
depends on the radial trend of this latter quantity which,
for the isotropic models here considered is given by
\begin{eqnarray}
\rhos\sigast^2 =\rhos\sigma^2_{*\circ}+
                {G\mast\mbh\over 4\pi\rs^4}
                    \left[{1-2s+6s^2+12s^3\over 3s^3(1+s)}
                  -4\ln\left(1+{1\over s}\right)\right],\quad 
                 s\equiv {r\over\rs},
\label{jeans}
\end{eqnarray}
where $\sigma_{*\circ}$ is the isotropic 1-dimensional stellar
velocity dispersion without the contribution of the SMBH (Ciotti et
al. 2008). Note that, at variance with the estimate of $\sigap$, in
the thermalization of the stellar mass losses we also consider the
(time increasing) contribution of the gravitational field of the
central SMBH.

\subsection{The circumnuclear disk and the SMBH accretion luminosity}

At the onset of the cooling catastrophe a large amount of gas suddenly
flows onto the central regions of the galaxy, and this induces star
formation and accretion on the central SMBH, producing a burst of
energy from the galaxy center.  However observations of our own
galactic center and high resolution studies of other nearby systems
indicate that, in addition to the central starburst with radius $\sim
100 - 300$ pc, accretion onto the central SMBH is mediated by a small
central gaseous disk within which additional significant star
formation occurs, and the remaining fraction of gas either is blown
out in a BLR wind, or it is accreted onto the central SMBH.  In our
treatment the disk is not simulated with hydrodynamical equations, but
its description is needed to obtain important quantities required by
the code.  In an improvement over CO07, we now also model and add to
the hydrodynamics the mechanical feedback produced by the disk wind
(see Table 1), but for future reference, in the following we also
describe the modeling of a nuclear jet. However, while the jet
contribution to the circumnuclear disk mass balance is considered in
the numerical integration, the associated mechanical feedback is not
added to the current hydrodynamical simulations.

The circumnuclear disk, which is the repository of the gas inflowing
at a rate $\dmineff$ from the first active mesh point $R_1$ of the
hydrodynamical grid, and which feeds the central SMBH at a rate
$\dot\mbh$, contains at any time the mass gas $\Mdg$ and a total
stellar mass $\Mds =\Mdsl +\Mdsh$, which is divided among low and high
mass stars (with the division mass at $8\Msun$).  The disk also
contains a mass $\Mrem$ of remnants from the earlier generations of
evolved stars.

In the adopted scheme the accretion rate on the central SMBH is given
by
\beq
\mdot ={\dMfid\over 1+\etaD},
\label{eq:mdotbh}
\eeq
where
\beq
\dMfid\equiv {\Mdg\over\tlagd},\quad 
\etaD\equiv {\dMfid\over 2\dot\Medd},\quad
\dot\Medd\equiv{\ledd\over\epsz c^2}
\label{eq:mfid}
\eeq 
are the fiducial depletion rate of gas from the circumnuclear disk,
its normalized value, and the Eddington mass accretion rate,
respectively. The reference radiative efficiency $\epsz$ is defined in
eq.~(\ref{eq:ADAF}).  Equations (\ref{eq:mdotbh})-(\ref{eq:mfid}) are
designed to guarantee that when $\etaD\ll 1$ the gas is accreted onto
the central SMBH at the rate $\dMfid$, while $\dot\mbh=2\dot\Medd$ for
$\etaD\gg 1$ (i.e., we allow for possible moderate super-Eddington
accretion; note however that outside the first grid point $R_1$ the
flow accretion rate is limited in a self-consistent way by feedback
effects). From eq.~(\ref{eq:mdotbh}) we calculate the instantaneous
bolometric accretion luminosity as
\beq
\lbh =\epsA\,\mdot\,c^2,
\label{eq:lbhbol}
\eeq
and, at variance with CO07 (see also Table 1), here we adopt an
``ADAF-like'' radiative efficiency
\beq
\epsA =\epsz{A\dot m\over 1+A\dot m},\quad 
\dot m\equiv {\mdot\over \dot\Medd},
\label{eq:ADAF}
\eeq
where $A$ is a free parameter so that $\epsA\sim \epsz A\dot m$ for
$\dot m \ll A^{-1}$.  In our simulations we fix $A=100$ (see,
e.g. Narayan \& Yi 1994, and CO01, where a very preliminary
investigation of ADAF effects on radiative feedback was carried out),
and we adopt $\epsz=0.1$ or 0.2 (e.g., see Noble, Krolik \& Hawley
2008).  As usual in accretion theory, we finally introduce the
normalized accretion luminosity
\beq
l\equiv{\lbh\over\ledd}={A\dot m^2\over 1 + A\dot m},
\label{eq:lbhnor}
\eeq
where the last expression derives from the ADAF phenomenological 
description.

There are a few {\it lag times} in our problem which are expressed as follows.
The first is the {\it instantaneous disk lag time}, appearing
in eq.~(\ref{eq:mfid}), and already considered in CO07, defined as
\beq
\tlagd\equiv{2\pi\over\alpha}\sqrt{\Rd^3\over G\mbh}, 
\label{eq:tlagd}
\eeq 
where $\alpha\simeq 10^{-2}-10^{-1}$ is the disk viscosity
coefficient, and $\Rd$ and $\mbh$ are the instantaneous values of the
fiducial radius of the circumnuclear disk and the mass of the central
SMBH.  In CO07, the disk radius $\Rd$ was maintained fixed to $R_1$,
instead now we use the scaling predicted by thin-disk theory
\beq
\Rd(t) =f_d R_1\times\left ({\mbh \over M_{\rm BH0}}\right)^{2/3},
\label{eq:rdisk}
\eeq 
(e.g., see Morgan et al. 2007) where $M_{\rm BH0}$ is the central SMBH
mass at the beginning of the simulation. We assume $f_d=0.4$, so that
$\Rd (0)\simeq 2$ pc for an initial SMBH mass of $\simeq 10^8\Msun$.

The second characteristic time (that was not considered in CO07) is the
instantaneous {\it infall} lag time from $R_1$ to the disk, estimated
as
\beq
\tlagi={R_1\over\vff},\quad 
\vff\equiv\sqrt{2G\mbh\over R_1},
\label{eq:tff}
\eeq
so that the effective rate at which gas accretes on the disk is obtained 
by solving the differential equation
\beq
{d\dmineff\over dt}={\dmin -\dmineff\over\tlagi},
\label{eq:dmineff}
\eeq
where $\dmin$ is the instantaneous rate at which gas flows through the
first active grid point\footnote{$\dmin$ is taken positive in case of
  accretion and zero in case of outflow at $R_1$.}.  It follows that
when $\dmin$ provided by hydrodynamics drops to zero the circumnuclear
disk experiences a fueling declining exponentially with time.

The disk total gas mass $\Mdg$ is not only the source of SMBH
accretion, but also of star formation in the disk: we assume that a
fraction of $\Mdg$ is converted into stars at a rate $\etas\dMfid$
(where $\etas\simeq 10\Mdg/\mbh$), and that another fraction of $\Mdg$
is lost as a {\it disk wind} and as a {\it jet} at instantaneous rates given by
$\etaw\dot\mbh$ and $\etaj\dot\mbh$, so that the equation for the gas
mass in the disk is
\beq
{d\Mdg\over dt}=\dmineff - (1+\etaw+\etaj)\dot\mbh -\etas\dMfid.
\label{eq:dmdg}
\eeq 
Note that the presence of a jet was not considered in CO07, so that
$\etaj =0$ therein.  The stars formed in the disk are described
separately as a function of their mass, i.e., high-mass stars ($M>\MII
=8\Msun$) produce a total disk mass $\Mdsh$, and low-mass stars ($\Min
<M<\MII$) contribute to a disk mass $\Mdsl$ according to the equations
\beq
{d\Mdsl\over dt}=(1-\fh)\etas\dMfid -{\Mdsl\over\taul};\quad
{d\Mdsh\over dt}=\fh\etas\dMfid -{\Mdsh\over\tauh}.
\label{eq:dmdstar}
\eeq
For the characteristic evolutionary times we adopt $\taul
=\tauopt$ and $\tauh=\tauII$ given in CO07, while we assume $\fh =
0.5$, corresponding to a top-heavy Salpeter-like initial mass function
(e.g., see Nayakshin \& Sunyaev 2005, Nayakshin et al. 2006) of
slope $x\simeq 1.16$ and minimum mass $\Min = 0.1\Msun$.  The
associated optical ($\ldopt$) and UV ($\lduv$) luminosities of the
stellar disk are calculated following the scheme described in CO07.
Finally stellar remnants mass in the disk evolves as
\beq
{d\Mrem\over dt}=\freml{\Mdsl\over\taul}+\fremh{\Mdsh\over\tauh},
\label{eq:dmdrem}
\eeq
where $\freml=0.2$, $\fremh=0.09$. 
\begin{figure}
\includegraphics[angle=0.,scale=0.9]{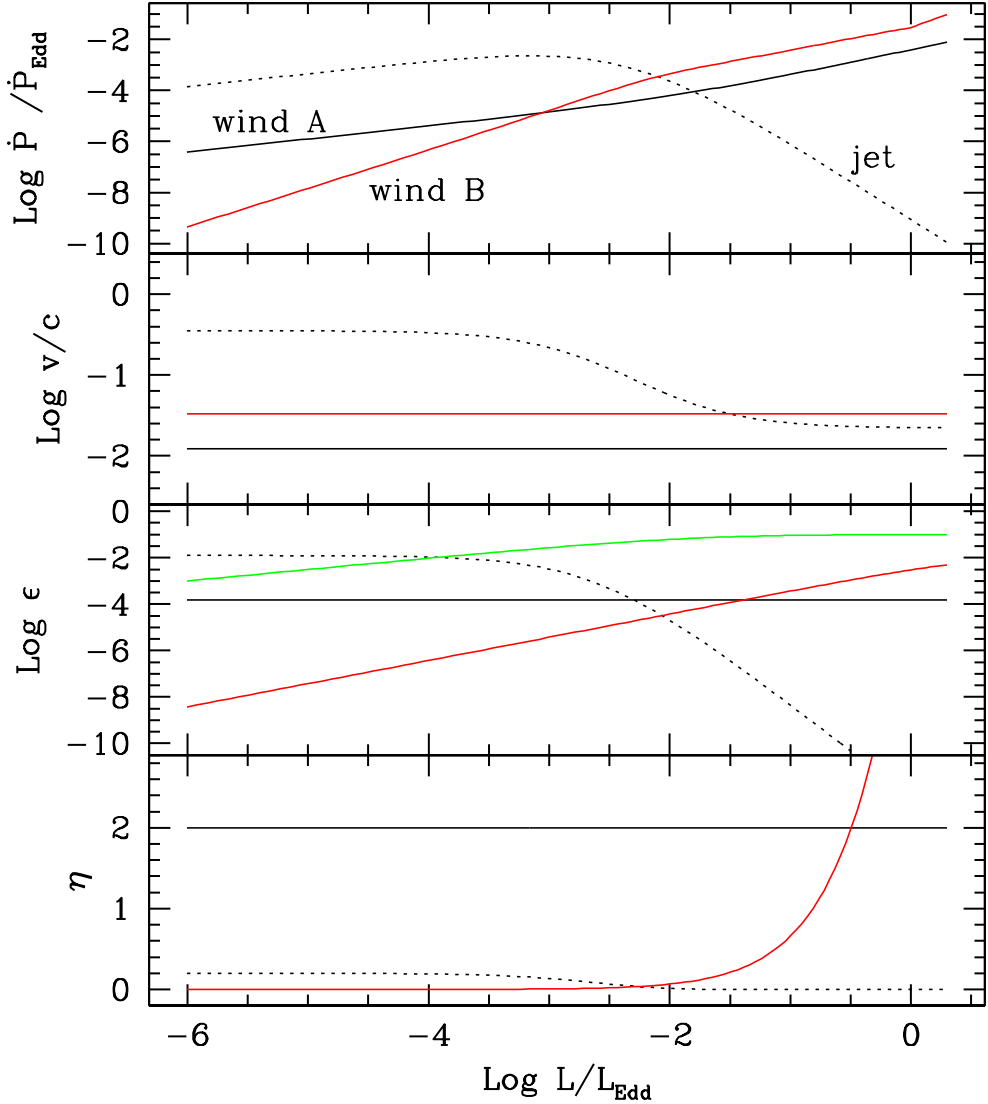}
\caption{Mechanical feedback properties as a function of Eddington
  ratio $l$ with ADAF coefficient $A=100$ (eq.~[\ref{eq:lbhnor}]).
  Red solid lines refer to wind properties of Type B models with
  maximum wind efficiency $\epswM=5\;10^{-3}$ and $\etawM=1800\epswM$
  (eq.~[\ref{eq:modB}]).  From bottom to top panels: ejected mass
  fraction (eqs.~[\ref{eq:etaw}]-[\ref{eq:modB}]); wind mechanical
  efficiency (eq.~[\ref{eq:epsw}]); wind velocity
  (eq.~[\ref{eq:vdiskw}]); normalized wind momentum per unit solid
  angle $\dot P/\dot P_{\rm Edd}=\sqrt{2\epsw\etaw}\dot m/(2\DOmew)$
  (where $\dot P_{\rm Edd}=\dot\Medd c$, see eq.~[\ref{eq:momwj}]).
  The wind opening angle is given in eq.~(\ref{eq:domeg}). Black solid
  lines refer to wind properties of Type A models, where $\epsw=
  1.5\;10^{-4}$, $\etaw=2$, $\DOmew=\pi$. Dotted lines represent the
  corresponding jet properties, parametrized as in eqs.~(\ref{eq:dmj})
  and (\ref{eq:ldjet})-(\ref{eq:momwj}), with
  $\DOmew=2.5\;10^{-2}$. Finally, the green line is the ADAF modulated
  EM efficiency as given by eq.~(\ref{eq:ADAF}), with $\epsz=0.1$ and
  $A=100$.}
\label{f1}
\end{figure}
The equation for the mass loss associated with the disk wind is 
\beq
{d\Mdw\over dt}=\etaw\mdot+
               (1-\freml){\Mdsl\over\taul}+
               (1-\fremh){\Mdsh\over\tauh}:
\label{eq:mdiskw}
\eeq
the first term is a mass loss driven as a wind by the central SMBH, and the
second and third are from high mass and low mass stars in the central
disk.

At variance with CO07, where the effects of mechanical feedback were
not considered and the wind treatment was extremely simplified, here
we explore two different classes of models, that we call {\it Type A}
and {\it Type B}. In particular, we use for the ratio of the wind
outflow rate to the SMBH accretion rate 
\beq
\etaw\equiv\cases{2,\quad\quad\quad\quad\quad\quad\quad\quad\,\; {\rm
    [A]}\cr \displaystyle{{3\etawM\over 4}{l\over 1+0.25 l}},
  \quad\quad\quad {\rm [B].}}
\label{eq:etaw}
\eeq
Therefore, Type A models correspond to the models in CO07.  In case of
Type B models, whose parameterization is introduced to mimic the
results of 2D simulations, the efficiency of ejecting a wind increases
with increasing Eddington ratio, $\etawM$ fixes the maximum value of
$\etaw$, and the factor $3/4$ takes into account the maximum possible
value for the scaled accretion luminosity $l$.  The situation is
illustrated in the bottom panel of Fig.~1, where the solid lines
represent the coefficient $\etaw$ for Type A models (black) and Type B
models (red, with $\etawM=1800\epswM$).  As a consequence of this
choice, winds in Type A models eject more mass at low accretion
luminosities, while the opposite happens in Type B models for $l\gsim 1/3$.

In particular, our scaling of the radiation driven wind mass loss rate
and mechanical power are motivated both by current simulations and by
observations of massive outflows from quasars (e.g., Ganguly \&
Brotherton 2007; Holt, Tadhunter \& Morganti 2008). Some quasars show
broad absorption lines (BALs) which are the most dramatic evidence for
winds in AGN. BALs are almost always blueshifted relative to the
emission-line rest frame, indicating the presence of outflows from the
active nucleus, with velocities as large as $0.2~c$ (e.g., Turnshek
1998).  BALs are observed not only in the UV but also in the
X-rays: for example, Chartas, Brandt \& Gallagher (2003) discovered a
very broad absorption line in the X-ray spectrum of PG~1115+80. It is
commonly accepted that these mass outflows in quasars are very likely
disk winds driven by radiation from accretion disks (e.g., see
K\"onigl 2006, Proga 2007a).

Of course, determining the mass loss rate and mechanical power based on
observations requires modelling, because observations do not directly
provide the information about the wind column density and
photoionization structure (e.g., Arav et al. 2007, and references
therein). In addition, one must assume something about the wind
covering factor. Therefore, a physical model of disk winds is needed
to complement data analysis and to estimate the key wind properties.
In particular, hydrodynamical simulation of radiation driven disk winds
allows us to explore the impact upon the mass-loss rate, $\dot\Mdw$ and
outflow geometry caused by varying the system luminosity and the
radiation field geometry. For example, Proga at
al. (1998) showed that winds driven from, and illuminated solely by,
an accretion disk yield {\it complex, unsteady outflow}.  In this
case, time-independent quantities can be determined only after
averaging over several flow timescales. On the other hand, if winds
are illuminated by radiation mainly from the central engine, then the
disk yields steady outflow. Proga et al. (1998, see also Proga 1999)
found that $\dot\Mdw$ is a strong function of the total luminosity,
while the outflow geometry is determined by the geometry of the
radiation field. In particular, for a relatively small $l$,
there is no wind but only a puffed up disk. Then as $l$
increases a strong equatorial wind develops. For very high
$l$, the wind becomes bipolar.  For high system luminosities
$\dot\Mdw$ of the disk scales with the luminosity in a way similar to
stellar mass loss rate, i.e., $\dot\Mdw\propto l^{1.7}$ (see Castor,
Abbott \& Klein 1975).  As the system luminosity drops below a
critical value [$\sim 2\ledd/(1+M_{\rm max})$], $\dot\Mdw$ decreases
quickly to zero ($M_{\rm max}$ is the total line opacity for an optically
thin case).

The wind power and geometry depend on the luminosity and also on the
mass of SMBH. Proga \& Kallman (2004, see also Proga et al. 2000)
showed that for $\mbh=10^8 \Msun$ and $l\simeq 0.5$ a strong wind
develops whereas for $l\simeq 0.1$ there is no disk wind (eq.~[18] for
Type B models reflects this trend). The primary reason for this
luminosity sensitivity is the fact that the mass flux density of the
wind decreases strongly with decreasing disk luminosity and the wind
is more subject to overionization.  Proga \& Kallman (2004) also found
that for a fixed Eddington fraction (e.g., 50\%) it is easier to
produce a wind for $\mbh > 10^7 \Msun$ than for $\mbh < 10^7\Msun$.
This result is a consequence of the decrease of the UV contribution to
the disk total radiation with decreasing mass of the BH in the Shakura
\& Sunyaev (1973) disk model for a fixed Eddington fraction.

In addition to the treatment of CO07, we now consider another mass 
component ejected by disk, i.e. a {\it nuclear jet} with instantaneous 
mass flow
\beq
{d\Mj\over dt}=\etaj\mdot,\quad\etaj={0.2\over (1+100l)^4};
\label{eq:dmj}
\eeq
the trend of $\etaj$ as a function of $l$ is shown in the bottom panel
of Fig.~1 with the dotted line: it is apparent how the mass ejected by
the jet is always negligible with respect to the wind mass loss in
Type A models, while it is slightly dominant over the wind in Type B models 
at low luminosity ratios.

We stress that, even though we consider this additional mass component
in the circumnuclear wind mass balance (eq.~[\ref{eq:dmdg}]), the
associated feedback effects on the galaxy ISM are not taken into
account in the simulations (see Table 1), and for the moment we are
assuming that the mass, momentum, and energy fluxes in the jet escape
the galaxy. In the code, all the equations presented in this Section
are integrated numerically with a first order finite difference
scheme.

\subsection{The mechanical feedback treatment}

We now discuss how the kinetic energy, momentum and mass of the BLR
wind are transferred to the ISM. For sake of completeness, we also
describe the jet treatment.

As in CO07, the fiducial {\it instantaneous mechanical luminosity} of the 
disk wind is given by
\beq
\ldwin =\epsw\dot\mbh c^2 +\epsII c^2 (1-\fremh){\Mdsh\over\tauh},
\label{eq:ldwin}
\eeq
where $\epsw$ is the mechanical efficiency of the wind, and the second
term describes the energetic associated with the SNII explosions of
the high-mass stars in the circumnuclear disk. In analogy with
eq.~(\ref{eq:etaw}), we assume
\beq
\epsw\equiv\cases{\epswM,\quad\quad\quad\quad\quad\quad\quad\;\;\, {\rm [A]}\cr
                  \displaystyle{{3\epswM\over 4}{l\over 1+0.25 l}},
\quad\quad\quad {\rm [B].}}
\label{eq:epsw}
\eeq
In Type A models we explore the range $3\,10^{-5}\leq\epswM\leq
5\,10^{-3}$ (see Table 2), i.e also values significantly lower than
adopted by Hernquist and collaborators ($5\,10^{-3}$, see Di Matteo et
al. 2005).  For reference, in CO07 (where $\epsw$ was also kept
constant), we adopted $\epswM =5\, 10^{-4}$, but the resulting mass,
energy and momentum sources were not added to the code.  In case of
Type B models, where the wind efficiency is a function of the
normalized accretion luminosity, $\epswM$ is the maximum possible
value (reached for $l=2$), and the different values adopted in the
simulations are listed in Table 2.  In both cases the {\it
  instantaneous disk wind velocity} is given by
\beq
\vw\equiv\sqrt{2\ldwin\over\dot\Mdw}\simeq\sqrt{2\epsw\over\etaw}c,
\label{eq:vdiskw}
\eeq
where the last expression neglects the mass return contribution of
massive stars in the circumnuclear disk (see eq.~[\ref{eq:ldwin}]).
Note that in Type A models $\vw$ is in the range $2\,10^3 - 2\,10^4$
km s$^{-1}$ (as a function of the specific assumed value for
$\epsw$), in agreement with observations of BLRs (e.g., Crenshaw et
al. 2003). For the same reasons, in Type B models we require
$\vw=10^4$ km s$^{-1}$, so that $\etawM$ and $\epswM$ are linked by
the relation
\beq
\etawM=1800\epswM.
\label{eq:modB}
\eeq

In analogy with the wind component, the {\it instantaneous jet 
mechanical luminosity} is written as 
\beq
\lj =\epsj\mdot c^2,\quad \epsj={0.0125\over (1+400l)^4},
\label{eq:ldjet}
\eeq
and the jet velocity is given by
\beq
\vj\equiv\sqrt{2\lj\over\dot\Mj}=\sqrt{2\epsj\over\etaj}\,c,
\label{eq:vdiskj}
\eeq
which, for our chosen parameterization gives high but subrelativistic
jet velocity of $\vj/c\simeq 10^{-1.65}$ for $l\gsim 0.1$ (see
Fig.~1). Note that from eqs.~(21) and (24) jets are far more efficient
than nuclear winds at low Eddington ratios (Ghisellini 2008, Jolley \&
Kuncic 2008).  The dependence of wind and jet velocities on the
normalized accretion luminosity is shown Fig.~1 (second panel from the
top), where solid lines refer to the wind and the dotted line to the
jet.  Finally, the wind and jet momentum are defined as
\beq
\mj\equiv\dot\Mj\vj;\quad \mw\equiv\dot\Mdw\vw. 
\label{eq:momwj}
\eeq
We now illustrate how we distribute the mechanical feedback over the
galaxy ISM. In the following we describe the procedure both for the
wind and for the jet, even though in the hydrodynamical equations we
take into account only the wind component.  First we introduce the
{\it instantaneous wind and jet lag times}
\beq
\tau_{\rm wj}\equiv{R_1\over \vwj}
\label{eq:tlagw}
\eeq 
from the center to the first active grid point $R_1$ (where the
subscript indicates the specific component - disk wind or nuclear jet
- considered), and at each time step we compute the time-lagged values
for mass, momentum, and kinetic energy at $R_1$ by solving the
differential equation
\beq
{dX_l\over dt}={X-X_l\over\tau_{\rm wj}},
\eeq
where $X_l$ is the generic lagged variable associated with the instantaneous
unlagged value $X$.  Outside $R_1$ we then distribute
mass, momentun and kinetic energy over the hydrodynamical grid 
(outside $R_1$), by integrating numerically the phenomenological differential
equation
\beq
{\partial\ln\Ywj\over \partial\ln r}=-{\Pism (r)\over\Pwj (r)}-
                            {r\over \vwj}
                            {\partial\ln\Ywj\over\partial t},
\label{eq:mechfed}
\eeq
where $\Ywj$ is the mass, momentum and energy of the disk wind/jet
component at distance $r$ from the center, $\Pwj(r)$ is the local
wind/jet pressure, and for each quantity $Y(R_1)=X_l$. In this paper
we restrict to simulations where the time derivative is neglected, 
while in Paper III also this latter term is considered. 
In practice, we first integrate eq.~(\ref{eq:mechfed})
for the wind/jet pressure, i.e.,
\beq
\Pwj={\Ywj\over 2\Delta\Omega_{\rm wj} r^2},
\label{eq:pwj}
\eeq
where $\Ywj$ is the effective wind/jet momentum crossing the shell of
radius $r$, so that eq.~(\ref{eq:mechfed}) is a non-linear
differential equation for $\Ywj$. Once the equation is integrated, the
radial behavior of $\Pwj$ and the r.h.s.  of eq.~(\ref{eq:mechfed})
are known over the whole grid, and the equation can be integrated for
mass and energy. We thus determine how much of the mechanical energy,
momentum and returned mass are deposited in the ISM at each radius.

A physical justification of eq.~(\ref{eq:mechfed}) is given in
Appendix A, but it can be thought of phenomenologically in the
following way. If the pressure corresponding to the momentum flow
within the jet or wind is much greater than the pressure in the
ambient gas, very little mass, momentum and kinetic energy is taken
from it and deposited in that ambient gas. But when the r.h.s. of
eq.~(\ref{eq:mechfed}) approaches unity, the ``working surface''
has been reached and the jet or wind discharges its content.

The solid angle in the denominator of eq.~(\ref{eq:pwj}) is the
opening angle of the wind and of the jet, and the factor of 2 accounts
for the biconical nature of the flow. While for the jet we
assumed in all the simulations the fiducial value
$\DOmej=2.5\;10^{-2}$, for the wind case we adopt
\beq
\DOmew=\cases{
\pi\quad\quad\quad\quad\quad\quad\quad\quad\quad\quad\;\;\;\,{\rm [A]}\cr
\pi\min(\sqrt{l^2+a^2},1),\quad\quad\quad\;{\rm [B]},}
\label{eq:domeg}
\eeq 
where case B is designed to mimic the behavior found in radiation
driven winds: higher luminosity corresponds to a larger opening
angles. The constant inside the square root is fixed to
$a=\DOmej/\pi$, so that for small values of accretion luminosity the
wind opening angle coincides with the jet opening angle. Finally, note
that the almost linear dependence of $\DOmew$ on $l$ for $l>a$ assumes
that the {\it linear} opening angle depends on $\sqrt{l}$ for this
regime.  The wind/jet momentum (with the opening solid angle factor)
as a function of $l$ is shown in Fig.~1 (top panel), and it is
apparent that this quantity is dominated by the wind contribution
(solid lines) for $l\gsim 10^{-2}$.

To implement numerically the mechanical feedback terms, we finally
compute the nuclear wind mass, momentum and kinetic energy per unit
volume deposited in each shell as
\beq
{\rm Source}_{\rm wj}={3\over 4\pi}{\Ywj (R_i)-\Ywj (R_{i+1})\over 
                       R_{i+1}^3-R_i^3}
\eeq
and we add them (only for the wind component) to the r.h.s.  of
eqs.~(56)-(58) in CO07.  In practice, we assume that the quantities of
mass, momentum and kinetic energy transferred to the ISM within the
opening angle are finally distributed over the whole solid angle, a
forced choice for our 1D code.

Summarizing, our simple formulae for Type B models capture some of the
key wind properties of quasar winds: the wind opening angle, mass loss
rate, and power increase with the accretion luminosity, and the fact
that there are negligible radiation pressure driven winds for $l\lsim
0.01$.  Figure 1 shows (going downwards from the top) our modeling of
the inputs of momentum, velocity, energy and mass, respectively. The
simplest case of constant wind mechanical efficiency (Type A) is shown
as a solid black line, whereas the one in better accord with
radiatively driven winds (Type B) is shown in red. A hypothetical
relativistic jet model (``radio mode'') is given by the dotted lines
and the green line indicates the ADAF-like treatment of the radiative
efficiency. For Eddington ratios above 0.01 the solution represents
the typically observed (and computed) "AGN mode" of high radiative
efficiency and low mechanical energy output efficiency, whereas at low
Eddington ratios ($l\lsim 10^{-1.5}$) the output shifts to radio mode
of low EM output and high mechanical energy efficiency in a
relativistic jet. As we will see most of the SMBH growth occurs in the
AGN mode and so the net (mass weighted) efficiencies are in agreement
with well known estimates (e.g., Soltan 1982, Yu \& Tremaine 2002,
Merloni \& Heinz 2008, Martinez-Sansigre \& Taylor 2008), but much of
the energy output occurs in the radio mode with important consequences
for the energy balance in clusters of galaxies (e.g., see Mc Namara \&
Nulsen 2007, and references therein).

\subsection{The unchanged input physics}

The stellar mass loss rate and the SNIa rate associated with the
initial stellar distribution are the main ingredients driving
evolution of the models. In the code the stellar mass losses -- the
source of {\it fuel} for the activity of the SMBH -- follow the
detailed prescriptions of the stellar evolution theory, and we use
exactly the same prescriptions as in CO07 (see Sects.~2.2 and 2.3
therein).

The radiative heating and cooling produced by the accretion luminosity
are numerically computed as in CO07 by using the Sazonov et al. (2005)
formulae, which describe the net heating/cooling rate per unit volume
$\dot{E}$ of a cosmic plasma in photoionization equilibrium with a
radiation field characterized by the average quasar Spectral Energy
Distribution derived by Sazonov et al. (2004, see also Sazonov et
al. 2007, 2008), whose associated spectral temperature is $\tx\simeq
2$ keV.  In particular, Compton heating and cooling, \brem losses,
line and continuum heating and cooling, are taken into account. Also
the star formation over the galaxy body, the radiation pressure due to
electron scattering, to phoionization, and finally to UV, optical and
infrared photons on dust, are treated as in CO07, where the derivation
and the numerical integration scheme of the radiative transport
equations is described in detail.  Finally, all the relevant
information about the numerical code and the hydrodynamical equations
can be found in CO01 and CO07.

\section{Model  evolution}

We now show the main properties of representative models characterized
by a stellar mass $\mast= 2.9\times 10^{11}\Msun$, a blue optical
luminosity $\lb= 5\,10^{10}L_{\rm B\odot}$ (corresponding to a stellar
mass-to-light ratio in the blue band of $\simeq 5.8$), a Fundamental
Plane effective radius $\re=6.9$ kpc, and a central aperture velocity
dispersion $\sigap=260$ km s$^{-1}$. The stellar distribution is
immersed in a dark matter halo so that the total mass density
distribution is proportional to $r^{-2}$, and an identical amount of
stellar and dark matter is contained within the stellar half-mass
radius.  The initial SMBH mass follows the present day Magorrian
relation, with $\mbh\simeq 10^{-3}\mast$, as it is believed that the
bulk of the SMBH mass is assembled during the process of galaxy
formation (e.g., Haiman, Ciotti \& Ostriker 2004; Sazonov et
al. 2005), but this process is not modeled in the present
simulations. Therefore, these models are not appropriate as initial
conditions for cosmological simulations, because their parameters are
fixed to reproduce early-type galaxies similar to those observed in
the local universe (at $z=0$), and also because we set outflow
boundary conditions at the galaxy outskirts ($\sim 250$ kpc): from
this point of view, the simulations represent an isolated elliptical
galaxy (consistently we are not considering the effects of possible
merging on the galaxy evolution).  A central cluster galaxy would have
more difficulty generating winds and would suffer from bursts of
cluster gas inflow.  We adopted this framework to adhere to the
standard approach followed in ``cooling-flow'' simulations. In future
explorations we will address in a more consistent way the problem of
the galaxy structural and dynamical modifications due to star
formation and mass redistribution over an Hubble time, and the
compatibility of the obtained galaxies with the present-day scaling
laws of elliptical galaxies.  We also stress that the models presented
are just a representative sample out of several tens of runs that have
been made, characterized by different choices of the input parameters
(often outside the currently accepted ranges).

The initial conditions for the ISM are represented by a very low
density gas at the local thermalization temperature.  The
establishment of such high-temperature gas phase at early cosmological
times is believed to be due to a ``phase-transition'' when, as a
consequence of star formation, the gas-to-stars mass ratio was of the
order of 10\% and the combined effect of SNIa explosions and AGN
feedback became effective in heating the gas and driving galactic
winds (e.g., Sazonov et al. 2005). Several theoretical arguments and
much empirical evidence, such as galaxy evolutionary models and the
metal content of the Intra Cluster Medium support this scenario (e.g.,
Renzini et al. 1993; OC05; Di Matteo et al. 2005). For the reasons
above, in the simulations here presented (as well as in all others
simulations not shown), we assume that the galaxy stellar component at
the beginning of the simulation is 2 Gyr old, and the simulations span
~12 Gyr, so that the cosmic time at the end of the simulations is ~14
Gyr.

Important quantities associated with the model evolution are the mass
(luminosity) accretion weighted EM and mechanical efficiencies,
defined in natural way as
\beq 
<\epsA >\equiv {\int \epsA\mdot dt \over\Delta\mbh };\quad 
<\epsw >\equiv {\int\epsw\mdot dt \over\Delta\mbh}
\eeq 
where $\Delta\mbh$ is the SMBH accreted mass over the time interval
considered. In case of Type B models we also compute the luminosity
weighted average fraction of the sphere covered by the wind, defined
as
\beq 
<\DOmew>\equiv {\int 2\DOmew \lbh dt \over 4\pi \int \lbh dt}.
\label{eq:domegav}
\eeq 
At variance with eq.~(33) we restrict the computation of the integrals
to high luminosity phases only, that we fiducially assume defined by
$\lbh>0.1\ledd$.  Following CO07, we finally compute for the various
luminosities the {\it duty cycle} over a period of time $\Delta t$
\beq
f_{\rm duty}\equiv
{(\int^t_{t-\Delta t} Ldt)^2\over \Delta t\int^t_{t-\Delta t} L^2dt};
\eeq
with this definition a square wave with fraction of time $f$ in the high state
would have $f_{\rm duty}=f$.

We summarize the properties of the models in Table 2. For
reference we list first the cooling-flow ``CF'' model: with no SMBH
feedback (but allowance for SN feedback of types Ia and II, and taking
into account the SMBH gravitational field), it allows the central SMBH
to grow to mass far greater than observed in any system: clearly AGN
feedback is needed.

\subsection{Purely radiative models}

We now discuss a couple of {\it purely radiative} models (Table 2,
models RB0 and \RBZzd), i.e. models in which the mechanical feedback
effects are computed (following the Type B description) but not added
to the hydrodynamical equations, and whose maximum radiative
efficiency is $\epsz=0.2$ (\RBZzd) or $\epsz=0.1$ (RB0), values which
bracket current observational estimates for the likely bolometric
radiative efficiency (e.g., Fig.~2 in Yu \& Tremaine 2002).  We begin
by noticing that we do not expect that these models will be as
satisfactory as the models presented in CO07, at least for what
concerns the final SMBH mass.  In fact, we see from the first line in
Table 1 that not only the dark matter halo is much more important at
large radii than in our previous works (so that global degassing is
more difficult), but also that in the central regions of the galaxy
the gas is more tigthly bound. This is because the stellar density
profile is now proportional to $r^{-2}$, so that the gravitational
field is stronger in the central regions; in addition, the mass return
from evolving stars is correspondingly more concentrated and the
radiative losses more important.  Finally, radiative feedback is
reduced at low accretion luminosities, due to the ADAF
prescription. This reduction has significant effects: in fact, it is
during the low-luminosity phases that the stellar mass losses build up
the galactic gaseous atmosphere responsible for the successive cooling
catastrophe episodes. A very low accretion luminosity during the
quiescent phases contributes less to the global energetic balance, the
subsoic outflow at large radii (mainly due to SNIa heating) is less
favoured, and the time interval between the catastrophes is reduced
(see also Fig.~8 in CO01).

\begin{figure}
\includegraphics[angle=0,scale=0.8]{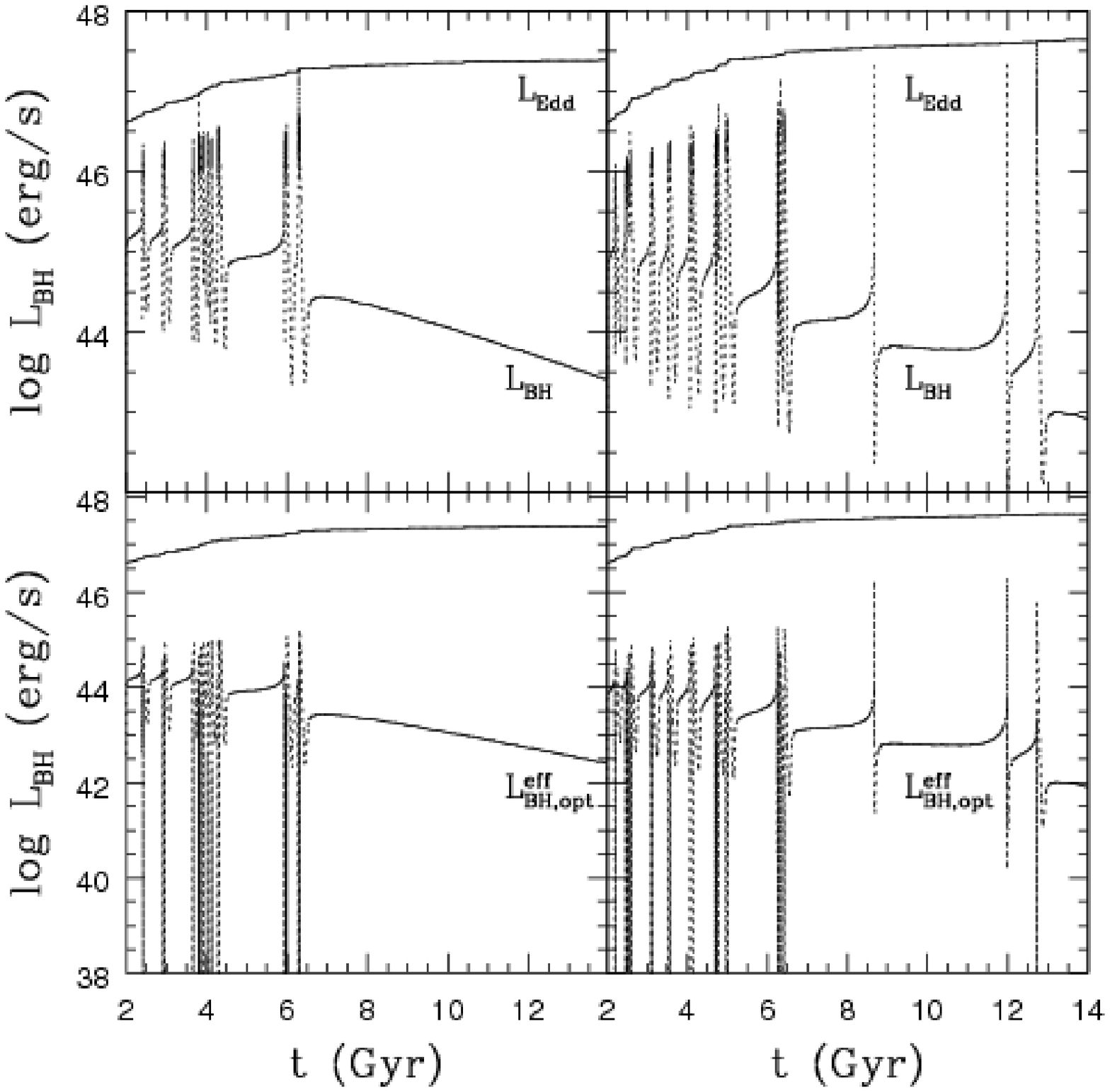}
\caption{Purely radiative models.  Dotted lines are the bolometric
  accretion luminosity (top) and the optical SMBH luminosity corrected
  for absorption, i.e., as it would be observed from infinity
  (bottom), for the models \RBZzd~(left panels), and RB0 (right
  panels). Model RB0 is identical to model \RBZzd, but with
  $\epsz=0.1$ instead of 0.2.  As in CO07 at the center we fixed
  $\lbhefopt(R_1)=0.1\lbh$. In model \RBZzd~quasar-like bursts at
  early times are followed at late times by a very quiet and passive
  accretion at late times with quite low optical luminosity.  With a
  lower radiative efficiency gas is ejected less efficiently from the
  galaxy and intermittent bursts persist to late times. While both
  models are far better than models without radiative feedback in
  addressing the problems caused by standard cooling flows, neither
  acceptably match the known properties of AGNs/young stars in massive
  galaxies (see Table 2).}
\label{fig:flr0}
\end{figure}

A first important result of the new models is that overall the main
properties of the CO01 and CO07 models are confirmed, and episodic
outbursts reaching $l\simeq 0.1$ are common.  After a first
evolutionary phase in which a galactic wind is sustained by the
combined heating of SNIa and thermalization of stellar velocity
dispersion, the central ``cooling catastrophe'' of the galaxy gaseous
halo commences, with the formation of a collapsing cold shell at $\sim
1$ kpc from the center.  In absence of the feedback from the central
SMBH a ``mini-inflow'' would then be established, with the flow
stagnation radius (i.e., the radius at which the flow velocity is
zero) of the order of a few hundred pc to a few kpc: these decoupled
flows are a specific feature of cuspy galaxy models with moderate SNIa
heating (Pellegrini \& Ciotti 1998). After the cooling catastrophe,
the radiative feedback caused radiation pressure and radiative heating
strongly affects the subsequent evolution, as can be seen in
Fig.~\ref{fig:flr0} where we show the luminosity evolution of the
central AGN with time-sampling of $10^5$ yrs.  The bolometric
luminosity (top panel) ranges between 0.001-0.1 of the Eddington limit
(the almost horizontal solid line) at peaks in the SMBH output but,
since obscuration is often significant, the optical accretion
luminosity as seen from infinity can be much lower (bottom panel): we
note that obscuration is commonly considered the explanation for red
quasars (e.g., see Wang 2008). However, the central quasar is not
always obscured and we see, in the lower panels of
Fig.~\ref{fig:flr0}, that the optical luminosity exceedes $\sim
10^{44}$ erg s$^{-1}$ in numerous bursts.  As already found in CO07,
the major AGN outbursts are separated by increasing intervals of time
(set by the cooling time and by the secular decrease of the mass
return rate from the evolving stellar population), and present a
characteristic temporal substructure, whose origin is due to the
cooperating effect of direct and reflected shock waves (from the inner
rim of the spherical strongly perturbed zone): {\it these outflowing
  shocks would be a likely place to produce emission of synchrotron
  radiation and electron and ionic cosmic rays}, that are considered
an additional source of feedback (Jiang, Ostriker \& Ciotti, Paper V,
in preparation; see also Sijacki et al. 2008).

In model \RBZzd~(left panels), at $t\simeq 6$ Gyrs the SNIa heating,
also sustained by a last strong AGN burst, becomes dominant and after
a last major burst a global galactic wind takes place and the nuclear
accretion switches to the optically thin regime.  The SMBH mass
accretion rate strongly oscillates as a consequence of radiative
feedback, with peaks of the order of 10 or (more) $\Msun$/yr, while
during the final, hot-accretion phase the almost stationary accretion
is $\lsim 10^{-2}\Msun$ yr$^{-1}$, much less than the istantaneous
mass return rate from the passively evolving stellar populations
($\dot\mast\lsim 1\Msun$ yr$^{-1}$). In the optical, $\log (\lbhefopt
/\ledd)\sim -5$ (see Table 2), consistent with observations (e.g., see
Ho 2008). However, the absolute accretion luminosity emitted ($\sim
3\,10^{43}$ erg s$^{-1}$) is still very large when compared to some of
the low-luminosity AGN observed (e.g., Pellegrini 2005a,b), showing that
some additional form of feedback is still needed in the low-luminosity
phase of the models. Note that in the last 6 Gyrs the SMBH virtually
stops its growth, while the ISM mass first increases due to the high
mass return rate of the evolving stellar population, and then
decreases due to the global galactic wind induced by SNIa.  During the
entire model evolution, a mass $\gsim 3\,10^{10}\Msun$ of recycled gas has
been added to the ISM from stellar mass losses.  Approximately
$2.1\times 10^{10}\Msun$ has been expelled as a galactic wind, while
$\sim 1.4\times 10^{10}\Msun$ has been transformed into new stars, so
that only 0.7\% of the recycled gas has been accreted onto the central
SMBH.  We note that several observational indications exist supporting
the idea that, while the majority of the stellar mass in elliptical
galaxies may have formed at high redshifts, small but detectable star
formation events (summing up to $\lsim 5-10\%$ of the total stellar
mass) may have occurred at low redshift (e.g., see Watabe, Kawakatu \&
Imanishi 2007; Pipino et al. 2008, Helmboldt, Walterbos \& Goto 2008;
Trager, Faber \& Dressler 2008).  Since the flow becomes nearly
steady, the duty-cycle approaches unity for RB02, but for RB0 the
short time in a strong burst at late epoch produces a small duty-cycle
as observed in low redshift central SMBHs.  Overall, the global
behavior of these purely radiative models is very similar to the
models in CO07, where we refer for details about the different
evolutionary phases.  Integrated values of their global quantities at
the end of the simulation are listed in Table 2.  Finally, it is
important to stress that an identical model without SMBH feedback
(i.e., $\epsz=0$ in eq.~[33]), but with the same star formation
treatment of the model just described, produced a SMBH of (the much
too high) final mass $\gsim 10^{10}\Msun$, while the total mass in new
stars was reduced to $\sim 3\times 10^9\Msun$ (see model CF in Table
2).  In addition, model CF does not present fluctuations in the
starburst and ISM X-ray luminosities, thus showing the vital
importance of SMBH feedback in the overall results.

Of course, reducing the value of the radiative efficiency from 0.2 to
0.1, the number of bursts and the final mass of the SMBH increases. In
particular, notice how major bursts persist to late times, (right
panels in Fig.~\ref{fig:flr0}): this behavior reduces considerably the
derived value of $f_{\rm duty}$ calculated at late times, and the
various quantities of the model are reported in Table 2.

Are these purely radiative models an adequate representation of
observed AGN/elliptical galaxies? While far better than cooling flow
models such as CF, which do not allow for radiative heating and
radiation pressure caused by the AGN, they are clearly inadequate in
several respects.  At the end of the simulations, the ratio of the
SMBH to stellar mass (also considering the added mass in new stars) is
$\sim 9\,10^{-3}$ (for model RB0) and $\sim 4.7\, 10^{-3}$ (for model
\RBZzd), larger than the canonical ratio of 0.0013 (e.g., Yu \& Tremaine
2002), i.e. radiative heating alone does not sufficie to
limit SMBH growth to the observationally allowed degree.  The
luminosity from young stars is $\sim 10^{40.2}$ erg/s even at
relatively low states between bursts, which would produce central
regions which are far too blue compared to typical ellipticals and the
AGN during these low states, while at only $10^{-4}\ledd$, is far more
luminous (at $\sim 10^{10.2}\Lsun$) than observed AGN in their low
states (see, e.g., Pellegrini 2005a,b, Wrobel, Terashima \& Ho 2007;
Gallo et al. 2007, Ho 2008).  Of course, from Table 2 it is apparent
that a model with a peak radiative efficiency of $\epsz=0.1$ is worst.

Are these defects due to too low an assumed radiative efficiency?  For
example, note that in model RB0 with a peak efficiency of 0.1, the
average (i.e., weighted over the mass accretion rate, see eq.~[33]) EM
efficiency of this model is seen to be 0.057 which is below that found
by Yu \& Tremaine (2002) by applying the Soltan (1982)
argument. However, model \RBZzd~ (with a peak radiative efficiency of
0.2), has an average radiative efficiency of 0.124, but is still
inadequate in representing real galaxies, especially at redshift
zero. In particular the relatively high values of the ratio
$\lbhefopt/\ledd$ combined with the final high values of $\mbh$
produces central optical luminosities $\gsim 10^{43}$ erg s$^{-1}$,
considerably above what is observed from nearby elliptical in the
``off'' state.  For these reasons we believe that the defects of the
purely radiative models cannot be ``fixed'' by simply increasing
$\epsz$; radiative feedback alone is insufficient (for massive
galaxies) in limiting cooling flow induced star formation and central
SMBH mass growth.

\subsection{Purely mechanical models}

\begin{figure}
\includegraphics[angle=0,scale=0.8]{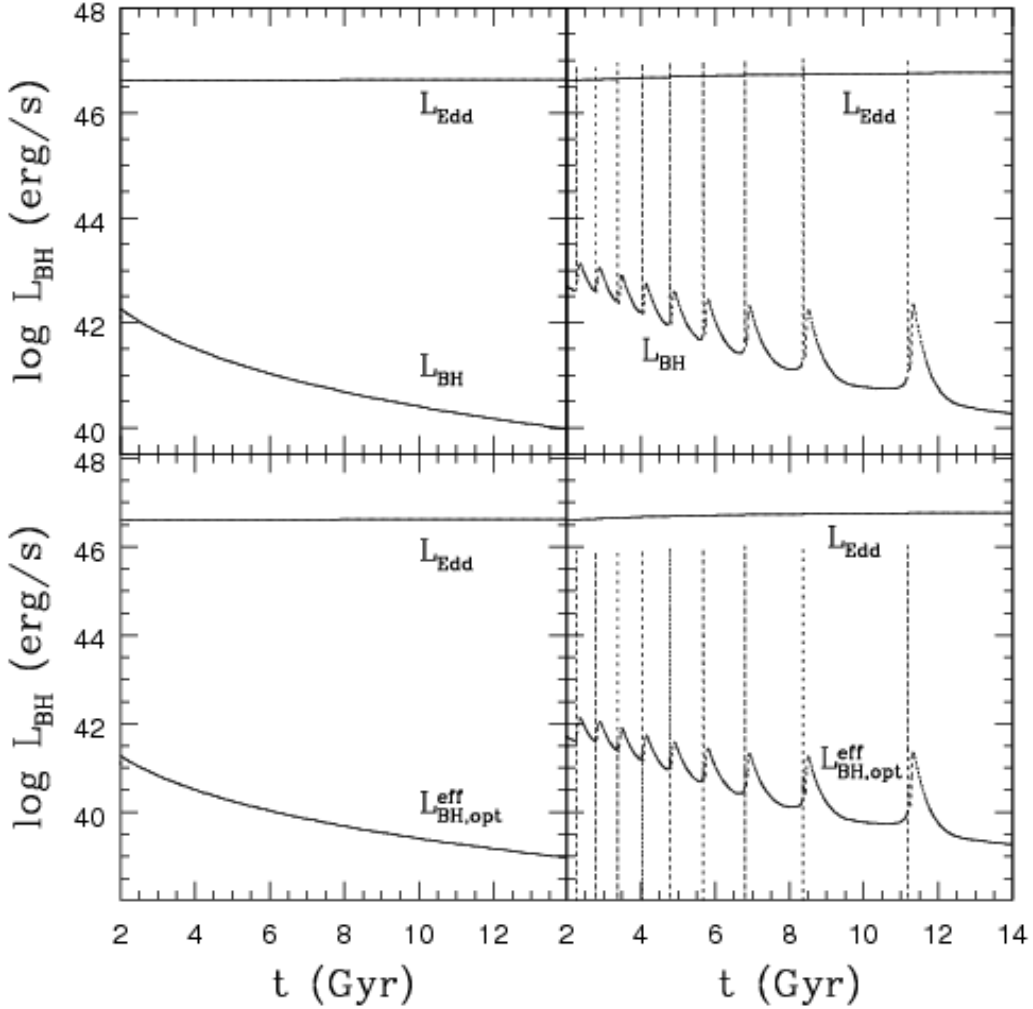}
\caption{Purely mechanical Type A models, i.e. models with a constant
  nuclear wind mechanical feedback efficiency.  Luminosity evolution
  of models MA0 ($\epswM=5\,10^{-3}$, left panels), and MA3
  ($\epswM=5\,10^{-5}$, right panels).  Dotted lines are the
  bolometric accretion luminosity (top) and the optical SMBH
  luminosity corrected for absorption, i.e., as it would be observed
  from infinity (bottom).  The global properties of Type A models are
  given in Table 2.}
\label{fig:flma}
\end{figure}
\begin{figure}
\includegraphics[angle=0,scale=0.8]{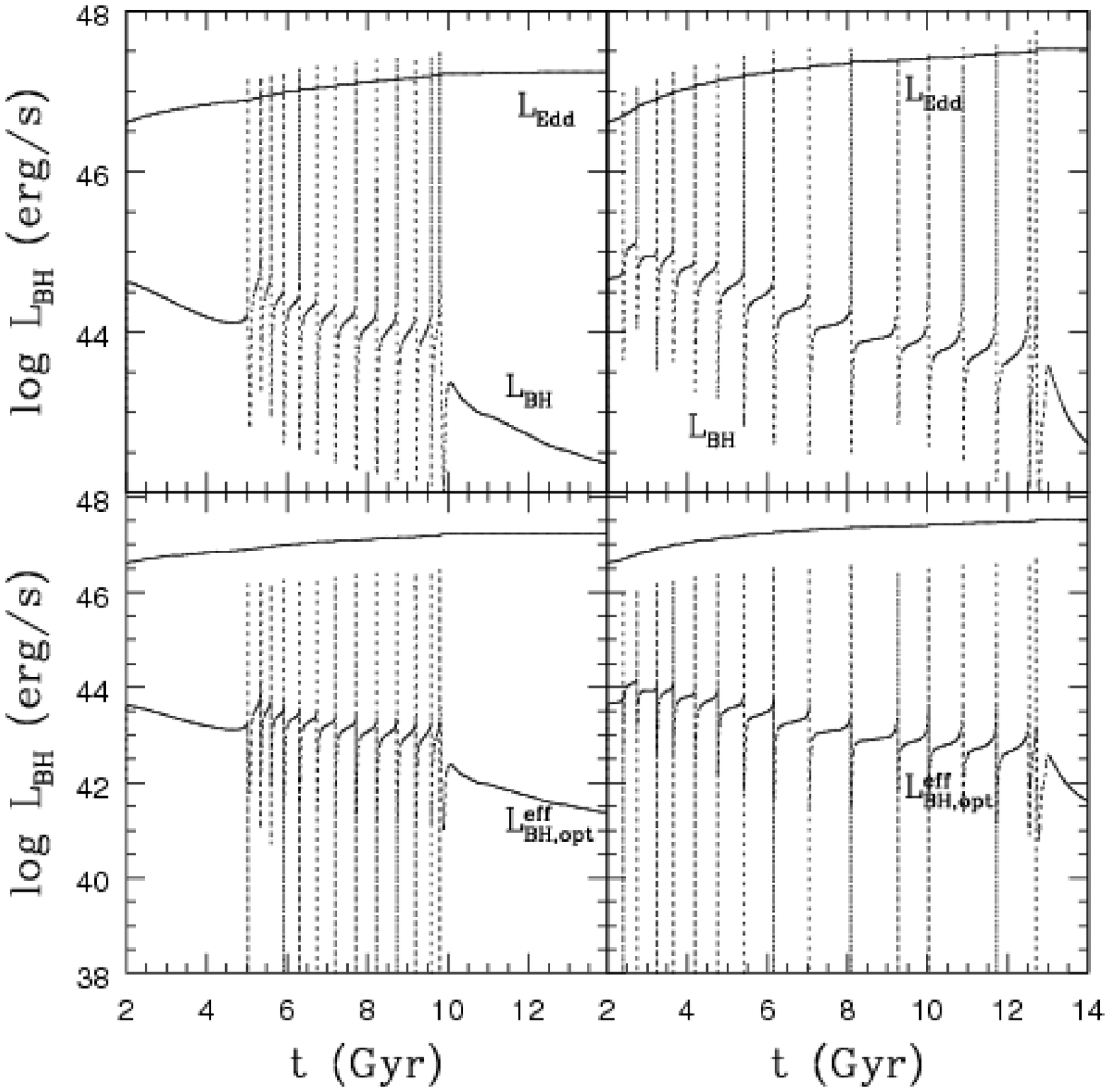}
\caption{Purely mechanical Type B models, i.e. models with a wind
  mechanical efficiency dependent on the (normalized) accretion
  luminosity $l$.  Luminosity evolution of models MB0 (the purely
  mechanical model associated with model RB0, left panels), and MB3
  (right panels). The nuclear wind mechanical feedback efficiency at
  peak are $\epswM=5\,10^{-3}$ and $\epswM=3\,10^{-4}$, respectively.
  Dotted lines are the bolometric accretion luminosity (top) and the
  optical SMBH luminosity corrected for absorption, i.e., as it would
  be observed from infinity (bottom).  The global properties of Type B
  models are given in Table 2.}
\label{fig:flmb}
\end{figure}

From the previous discussion, it follows that mechanical energy input
is also required.  In order to study the effects of mechanical
feedback we now switch off radiation feedback (while radiative cooling
is obviously maintained active), and we explore {\it purely
  mechanical} models.  In some sense, these models are complementary
to models RB0 and \RBZzd.  In a first assay, we add mechanical
feedback in the simple fashion normally adopted in several
investigations, as a fixed numerical efficiency for the input of
mechanical energy and for the amount of gas carried out by the wind.
These purely mechanical Type A models are referred in Table 2 as MA.
We then present the purely mechanical models of Type B (MB models in
Table 2), i.e. models in which the wind efficiency from the
circumnuclear regions depends on the accretion luminosity, declining
with declining values of $l$, as is expected for radiatively driven
winds. In both families, the model identification number in Table 2
increases at decreasing wind efficiency. In this paper we describe the
general properties of the purely mechanical models, while in Paper II
we focus on more specific quantitative aspects of purely mechanical
Type B models.

\subsubsection{Type A models: constant mechanical efficiency}

In these models the constant wind mechanical efficiency ranges from
$\epswM=5\,10^{-3}$ in model MA0, as utilized for example by Di Matteo
et al. (2005) and Johansson, Naab \& Burkert (2008), down to
$\epswM=3\,10^{-5}$ in model MA4.  The resulting global properties of
the whole family are reported in Table 2.

Model MA0 (Fig.~\ref{fig:flma}, left panels), with the highest assumed
steady efficiency is found in a state of permanent, global wind.  The
fact that the accretion luminosity is very low during the entire
evolution is revealed by the very low averaged value of
$<\epsA>\simeq 2.7\,10^{-3}$ obtained from eq.~(33), a consequence of
the ADAF implementation. With the assumed high mechanical efficiency we
find that (consistent with other investigators) almost all the
recycled gas produced by stellar evolution is ejected from the galaxy,
and a very negligible amount of mass is added to the central SMBH. In
addition, also star formation is maintained at very low levels. The
final X-ray thermal luminosity of the model is far too low, orders of
magnitude below that seen from normal giant elliptical galaxies (e.g.,
O'Sullivan et al.~2003).

As we reduce $\epswM$ (models MA1 and MA2), the evolution becomes more
interesting. In fact, gas accumulates via the usual thermal
instability, falls to the center and there generates an enormous
explosion and, in a single burst, the galaxy is essentially evacuated of
gas. There are no further bursts and the system is in permanent wind
state until $z=0$. This is essentially what is found by Di Matteo et
al. (2005) with the only difference being that the gigantic outburst
in their case is fueled by a gas rich merger while in our case it is
fed by recycled gas deposited at the center via the classic Field
thermal instability.  Since all of these models (MA0, MA1, and MA2)
have no bursts at late times and are ``on'' at a nearly constant but
low level, the corresponding duty-cycles at late times are near unity;
again, these models are too faint as thermal X-ray sources.

In model MA3 (Fig.~\ref{fig:flma}, right panels), the mechanical
feedback is quite weak, and the galaxy presents a series of bursts,
well separated by an increasing amount of time. The final SMBH mass is
almost doubled, and also the new stellar mass is now significant,
summing up to several $10^9\Msun$. The situation is even more extreme
in model MA4. These models match observations in one respect. In fact,
the occurrence of numerous bursts at late times is revealed by small
duty-cycles ($10^{-3}-10^{-4}$) so that they would be seen as AGN at
late times with approximately that probability, and this is roughly
consistent with observations. An inspection of Table 2 also reveals
that AGN feedback can be important not only to suppress star formation
when co-operating with SNIa to remove the gas from the galaxy, but
also to induce star formation, especially in the central regions of
the galaxies. This finding was already obtained in CO07, and proposed
as a possible way, in alternative or in addition to galaxy merging
(e.g., Hopkins et al.~2008a,b), to produce the central ``extra-light''
observed in some elliptical galaxy (see Lauer et al.~2005).

Clearly, these model cannot represent the cosmological evolution of
galaxies, as all galaxies would be found systematically at very low
values of their ISM X-ray luminosity, just consistent with discrete
sources, but in net disagreement with observations if considering high
efficiency models. In addition, a reduction of the fixed mechanical
efficiency leads to a quite strong transition to the situation in
which a significant amount of mass is added to the central SMBH. In
other words, purely mechanical models with constant wind efficiency
appear to be quite ``stiff'' in their behavior, switching from a
strong global wind to a recurrent and significant bursting activity
when reducing the efficiency.  Extreme ``fine tuning'' would be
required to produce an acceptable model and that result would probably
not be robust to the normal statistical variations of parameters
describing ellipticals in the Fundamental Plane. Due to the
computational time required, such statistical investigation is however
not attempted here, and it is postponed to Paper II.

Guided by these results, we reduce the importance of feedback by
considering a more plausible variation of the purely mechanical models
(Type B models), in which the mechanical efficiency is a function of
accretion luminosity, so that during bursts $\epsw$ is moderate, and
drops to very low values in the low-luminosity phases between bursts,
in better accord with theoretical models for radiative accretion of
AGN winds.

\subsubsection{Type B models: luminosity dependent mechanical efficiency}

We now move to describe the results for the family of MB models.  As
in the case of MA models, the value of the (peak) mechanical
efficiency decreases from model MB0 to model MB4. In particular, the
peak values $\epswM$ are coincident with the constant values of
$\epsw$ for the pairs MA0-MB0 and MA2-MB4, so that in these models the
effects of the luminosity-dependent mechanical feedback efficiency can
be studied by direct comparison.

In general, the effect of the luminosity dependence of $\epsw$ in Type
B models is an increase in the final mass added to the central SMBH,
as can be seen from Table 2. For example, at the end of the
simulations $\Delta\mbh\sim 10^{7.17}\Msun$ in model MA0, while
$\Delta\mbh=10^{8.98}\Msun$ in model MB0. Similarly, this is $\sim
10^{7.75}\Msun$ in model MA2 and $10^{9.44}\Msun$ in model MB4. Note
that these differences are mainly produced during the low-luminosity
phases, as at the peaks the mechanical efficiency is identical to the
corresponding MA models, and the comparison of the luminosity
evolution in the right panels of Fig.~\ref{fig:flma} and
Fig.~\ref{fig:flmb} illustrates this difference.  

As an additional example, in the right panels of Fig.~\ref{fig:flmb}
one can see a large number of bursts during all the evolution of model
MB3 ($\epswM=3\,10^{-4}$), while model MA1 ($\epswM=2.5\,10^{-4}$)
just presents a single burst.  And the accreted mass of the central
SMBH ($\Delta\mbh$) is in far better agreement with observations
(cf. MB2, MB3) than it is in the models with constant accretion
efficiency.  The present-day luminosity ratios of the (optical)
accretion luminosity to the Eddington luminosity are much smaller than
in the purely radiative models, but higher than in Type A models. The
duty-cycles (Table 2, column 13) are in general quite small, due to
the recurrent bursting activity which is still present in the models
at late epochs. It is interesting that the duration of a single burst
is of the order of $10^6$ yrs, in accordance with observational
estimates based on proximity effect (e.g., Kirkman \& Tyler 2008); we
also note how the luminosity averaged EM efficiencies, both in Type A
and Type B models, are in the range $0.04\lsim<\epsA>\lsim 0.08$, in
nice agreement with observational studies.

The amount of new stars formed is of the same order as in Type A
models, while the final mass of the ISM (and its X-ray luminosity) is
larger, due to the less efficient feedback during the low luminosity
phases. Also, the final SMBH masses are larger than those that are observed,
and larger than in Type A models. Therefore, while in Type A
models almost no mass is added to the central SMBH after the
process of galaxy formation (and so no cosmological evolution in the
$\mbh-\sigma$ relation is expected in absence of merging, in Type B
models a mass comparable to the initial SMBH mass is added on a
cosmological time, and a consequent evolution of the $\mbh-\sigma$
relation is expected. We note that observations are now becoming able
to probe such evolution (e.g., see Woo et al. 2008).  Finally, during
the low-luminosity phases, the nuclear emission is larger than what is
observed in (very) low-luminosity AGNs, and this is due to the reduced
contribution of the luminosity-dependent mechanical wind efficiency in
Type B models.

We then conclude that an additional form of feedback is needed both
during the high luminosity phases (and this feedback is in the form of
radiation, explored in Paper III, which is devoted to the study of
combined models), and during the quiescent low-luminosity phases at
late times. Obviously, this second form of feedback cannot be provided
by radiation pressure and radiative heating as in the peak phases, and
it is presumably provided by jets and/or thermally driven nuclear
winds, not included in the present models.  In sum, Type B models
(ADAF-like) are more satisfactory than the normally computed (fixed
efficiency) Type A models, but are still observationally inadequate.

\begin{deluxetable}{lccccccccccccc}
\rotate
\tablecaption{Properties of computed models}
\tabletypesize{\scriptsize}
\tablewidth{0pt}
\tablehead{
\colhead{Model}&
\colhead{$\epswM$}&
\colhead{$<\epsw>$}&
\colhead{\tablenotemark{a}$<\epsj>$}&
\colhead{\tablenotemark{a}$<\eps_{\rm EM}>$}&
\colhead{$\log \Delta\mbh$}&
\colhead{$\log \Delta M_*$} &
\colhead{$\log \Delta M_{\rm w}$}&
\colhead{$\log \mgas$} &
\colhead{$\log \lbhefopt/\ledd$} &
\colhead{$\log L_{\rm X,ISM} $}&
\colhead{$<\DOmew>$} &
\colhead{$f_{\rm duty}$} &
\\
\colhead{(1)}&
\colhead{(2)}&
\colhead{(3)}&
\colhead{(4)}& 
\colhead{(5)}&
\colhead{(6)}&
\colhead{(7)}&
\colhead{(8)}&
\colhead{(9)}&
\colhead{(10)}&
\colhead{(11)}&
\colhead{(12)} &
\colhead{(13)} &
}
\startdata
{\bf CF}\tablenotemark{b}   & ---     & ---   &  ---  & --- &10.34& 9.51 & 9.96  & 9.66 & ---& 41.31 & --- & ---\\
\hline
{\bf \RBZzd}   
            &\tablenotemark{a}$5\,10^{-3}$ &\tablenotemark{a}$1.7\,10^{-4}$ &$1.5\,10^{-3}$&0.124 &9.16& 10.22& 10.25& 9.73 & -5.08& 40.38 &0.094 & 0.6 0.7\\

{\bf RB0}  &\tablenotemark{a}$5\,10^{-3}$ &\tablenotemark{a}$1.8\,10^{-4}$ &$2.4\,10^{-3}$& 0.057 &9.45& 10.36& 10.29 & 9.69 & -5.72& 40.08& 0.098& 0.38 $8.2\,10^{-3}$\\
\hline
{\bf MA0}  &$5\,10^{-3}$   &$5\,10^{-3}$    &$1.2\,10^{-2}$  &$2.7\,10^{-3}$ &7.17  &6.43  &10.38 &7.65 &-7.71 &36.64 & --- & 0.79 0.79\\
{\bf MA1}  &$2.5\,10^{-4}$ &$2.5\,10^{-4}$  &$7.6\,10^{-3}$  &0.038         &7.56  &9.21  &10.41 &7.99 &-7.55 &37.60  & 0.50 & 0.79 0.73 \\
{\bf MA2}  &$10^{-4}$      &$10^{-4}$       &$6.0\,10^{-3}$  &0.051         &7.75  &9.46  &10.30 &9.75 &-7.52 &40.32  & 0.50 & $3.9\,10^{-4}$ 0.72\\
{\bf MA3}  &$5\,10^{-5}$   &$5\,10^{-5}$    &$3.7\,10^{-3}$  &0.069         &8.10  &9.72  &10.26 &9.88 &-7.55 &40.76  & 0.50 & $10^{-3}$ $1.7\,10^{-4}$\\
{\bf MA4}  &$3\,10^{-5}$   &$3\,10^{-5}$    &$2.0\,10^{-3}$  &0.073         &8.55  &9.83  &10.50 &8.04 &-7.76 &37.81  & 0.50 & $2\,10^{-3}$ $1.7\,10^{-3}$\\
\hline
{\bf MB0}  &$5\,10^{-3}$   &$8.8\,10^{-4}$   &$2.2\,10^{-3}$ &0.049         &8.98  &9.63  &10.24 &9.50 &-5.88& 39.91  & 0.46 & 0.92 $1.1\,10^{-3}$\\
{\bf MB1}  &$2.5\,10^{-3}$ &$5.4\,10^{-4}$   &$2.2\,10^{-3}$ &0.053         &9.11  &9.68  &10.28 &9.27 &-6.06& 39.52  & 0.46 & 0.96 $1.1\,10^{-3}$\\
{\bf MB2}  &$10^{-3}$      &$2.3\,10^{-4}$   &$2.4\,10^{-3}$ &0.056         &9.22  &9.67  &10.29 &9.42 &-6.04& 39.72  & 0.45 & $3.7\,10^{-2}$ $1.3\,10^{-3}$\\
{\bf MB3}  &$3\,10^{-4}$   &$7.8\,10^{-5}$   &$2.2\,10^{-3}$ &0.059         &9.33  &9.62  &10.30 &9.29 &-5.92& 39.49  & 0.43 & $3.9\,10^{-2}$ $1.9\,10^{-3}$\\
{\bf MB4}  &$10^{-4}$      &$2.4\,10^{-5}$   &$2.3\,10^{-3}$ &0.060         &9.44  &9.66  &10.30 &9.38 &-5.46& 39.41  & 0.43 & $3.9\,10^{-2}$ $2\,10^{-3}$\\
\hline
\enddata

\tablecomments{Masses are in units of Solar Masses and luminosities in
  $erg/s$. In models of class A the wind efficiency is maintained
  constant, i.e., $\epsw=\epswM$. In models of class B the value of
  $\epswM$ is reached when $\lbh=2\ledd$. In models with the subscript
  $02$ the maximum radiative efficiency in eq.~(8) is set to
  $\epsz=0.2$ instead of 0.1 as in the other models.  Mean
  efficiencies are calculated according to equation
  (34). $\Delta\mast$ is the total amount of star formed during the
  model evolution, $\Delta M_{\rm w}$ is the total amount of ISM lost
  at 10$\re$ and $\mgas$ the instantaneous amount of gas inside
  10$\re$. $\lbhefopt$ is the fiducial SMBH luminosity in the optical
  as would be seen at infinity after absorption, with
  $\lbhefopt=0.1\lbh$ at the first grid point (see CO07 for details).
  The luminosity weighted wind solid opening angle $<\DOmew>$ (in
  units of $4\pi$) is calculated according to eq.~(\ref{eq:domegav}),
  and the reported value refers to the final time of each simulation;
  in model MA0 the value is not reported as $\lbh < 0.1\ledd$ over all
  the evolution. The duty cycle is calculated for the effective
  optical accretion luminosity as would be seen from infinity.  The
  first number correspond to a Universe age of 4 Gyr, the second to
  13.7 Gyr, and the time span used for the computation is $\Delta
  t=t/2$.}
\tablenotetext{a}{Figures corresponding to quantities calculated but
  not added to the hydrodynamical equations.}
\tablenotetext{b}{The cooling flow model CF was stopped at 8 Gyrs.}
\end{deluxetable} 

\section{Discussion and conclusions}

In this paper we have investigated, with the aid of 1-D hydrodynamical
simulations with unprecedented spatial and time resolution (where the
cooling and heating functions include photoionization and Compton
effects, and restricting to models which include only radiative or
only mechanical feedback in the form of nuclear winds), the effects of
radiative and mechanical feedback on the gas flows in isolated
elliptical galaxies.  In particular, after having explored a few
purely radiative feedback models, we explored two sequences of models
in which the feedback is purely mechanical, and the only radiative
effects allowed in the simulations are those due to gas cooling.  The
first sequence (Type A models) has a fixed mechanical efficiency as in
most prior investigations, and the second (Type B models) allows the
efficiency to increase with increasing accretion rate as indicated by
the radiatively driven outflow models.  It is difficult to know,
either from fundamental physics calculations or from existing
observations, what values of the mechanical energy efficiency to adopt,
so we have tried a range of values from $5\, 10^{-3}$ to $10^{-5}$ to
see which, if any, provides acceptable results, consistent with what
is known about the properties of elliptical galaxies.

The investigation is in line with that of our previous papers, and the
general framework is maintained. However, the galaxy models and some
of the relevant input physics have been substantially improved and
extended (see Table 1).  As before, the recycled gas, arising from the
evolving stars in the inner several kpc of the galaxy (assumed a giant
elliptical), necessarily drives a classical radiative instability and
a collapse towards the center of metal rich gas. As a consequence, a
star-burst occurs and the central SMBH is fed.  We confirm that steady
accretion on SMBHs is only possible at very low Eddington ratios, and
consistently with the work of other authors we note that independent
of whether the feedback is mechanical or radiative no steady flow
appears to be possible for Eddington ratios above $\simeq 0.01$, and
that whenever the luminosity is significantly above this limit both
the accretion and the output luminosity is in burst mode.  The details
of how much gas is accreted on the central SMBH vs. consumed in stars
vs. ejected from the center by energy input from the starburst and
AGN, are one of the main outcomes of these models. Relevant
quantitative properties of the models are presented in Table 2, while
the general results can be summarized as follows:

1) Radiative heating and radiation pressure on the ISM by photons
emitted by the central AGN and by the starburst, without any
mechanical input, greatly reduces the ``cooling flow catastrophe''
problem, but leads to results that are still defective as compared to
detailed observations of local elliptical galaxies, in that the
central SMBH would be too bright and too massive, and the galaxy would
be too blue, due to repeated bursts of central star formation.

2) Adding mechanical energy from an AGN wind with fixed efficiency can
address the problems posed by pure cooling flows but does not give a
solution that in detail satisfies the observations. If the chosen
efficiency is large, then a giant burst and an explosive degassing of
the galaxy occurs (consistent with the results of Di Matteo et
al. 2005, Johansson et al. 2008, and other investigators).  Only a
small growth of the black hole occurs before the gas content of the
galaxy drops to levels below what is observed in real elliptical
galaxies and the systems at redshift $z=0$ are computed to have
thermal bremstrahlung X-ray luminosities orders of magnitude lower
than those typically seen in nearby ellipticals. Finally, we do not
get any late time AGN, and the duty cycle approaches unity. By
contrast, very low values of the assumed efficiency do not prevent the
cooling flows from causing more late time star formation (blue stellar
cores) than is seen in most nearby ellipticals. It appears that there
is no range of intermediate efficiencies that allows one to escape
from both of these defects.

3) Models with mechanical energy efficiency proportional to the
luminosity, as indicated both by observations and detailed 2-D
hydrodynamical simulations for radiatively driven winds, perform
better, but are still inadequate. We thus conclude that mechanical
energy input of the type adopted here - by itself - is unable to
provide appropriate levels of feedback that would leave ellipticals at
the current epoch with the properties that they are observed to have.
In fact, we find clear evidence that two additional different forms of
feedback are needed. The first, taking place during the high
luminosity phases, it is in the form of radiative feedback, and is
required to limit the growth of the central SMBH. The other form of
missing feedback is required during the quiescent, low-luminosity
accretion phases (in particular at late epochs), as revealed by
accretion luminosity values that, altough very sub-Eddington, are
still larger than observed. Of course, standard radiative feedback is
not effective during such phases, and presumably the further reduction
is provided by nuclear jets and/or thermally driven winds. We remark
again that these results are based on 1D simulations. It would be very
important to have similar 2D or 3D simulations, with the same input
physics, for a verification of the conclusions reported above, in
particular when considering the issues related to stronlgy aspherical
phenemona (such as the effective coupling of jets and nuclear winds
with the ambient ISM), or the effect of the ISM angular momentum on
the accretion rate.

4) We stress that in our study the initial conditions represent a galaxy
with a central SMBH closely following the Magorrian relation, and with
a stellar population already formed: such a system would be called a
``dead and red'' galaxy, without much evolution expected.  However,
one of the general consequences of our exploration is the fact that
the recycled gas from dying stars is a major source of fuel for the
central SMBH, and this can induce substantial QSO activity, {\it even
  in the absence of external phenomena such as galaxy merging}, and the
simulations reveal {\it how complex the evolution of an isolated
  galaxy, subject to internal evolution only, can be}.  We note that
recently similar conclusions have been reached, by using different
considerations, also by other authors (e.g., see Pierce et al. 2007, Li
et al. 2008, Kaufmann \& Heckman 2008).

Summarizing, neither models with purely radiative feedback nor those
with mechanical feedback as adopted here alone seem to be able to match
all of the basic observations of elliptical galaxies (see also Sijacki
et al. 2008).  Models with purely radiative feedback are unable, for
any reasonable radiative efficiency, to forestall sufficiently the
collapse of gas onto the central black hole. While they do reduce the
mass far below what it would have been in the no feedback ("cooling
flow") case, they do not reproduce the observed Magorrian relation and
leave the central black hole with masses greater than $10^9\Msun$,
perhaps a factor of four too large. The models with purely mechanical
feedback can, however, for an appropriate choice of the mechanical
efficiency, match the well established $\mbh-\sigma$ relation, but
only for efficiencies chosen to be so large as to leave the galaxies
with much less gas than is actually seen in normal ellipticals as
determined by their X-ray luminosities.  A more detailed discussion of
other aspects of the difficulties encountered by the models with
solely mechanical feedback, the description of the most successful
combined models (which utilize both radiative and mechanical
feedback), and their observational properties, is kept for the
following papers (Paper II, III, IV, and V).

\acknowledgments

We thank Jenny Green, Jeremy Goodman, Silvia Pellegrini, Eve Ostriker
and Jim Stone for useful discussions, and Annibale D'Ercole and Dennis
McRitchie for helpful suggestions on coding issues. We also thank the
anonymous Referee for a careful reading and for comments that improved
the presentation of the paper. D.P. acknowledges support by the
National Aeronautics and Space Administration under Grant/Cooperative
Agreement No.~NNX08AE57A issued by the Nevada NASA EPSCoR program.

\begin{appendix}

\section{A physical basis for the phenomenological
 mechanical feedback  equation: input from AGN wind/jet to 
hydrodynamic flow}

In this Appendix we describe the physics behind the phenomenological
differential equation (\ref{eq:mechfed}) modeling the wind and jet
mechanical interaction with the ISM of the galaxy: note that the
treatment is formally identical for the jet and for the wind.

How can we estimate the rate of energy deposition from a conical wind
or jet emitted by a central AGN into the ISM of the galaxy in which
the SMBH is embedded?  We will take from Sects.~2.2 and 2.3 the
estimates for the total outflowing mass rate in the wind or jet
($\dot\Mdw$ and $\dot\Mj$, eqs.~[\ref{eq:mdiskw}] and~[\ref{eq:dmj}];
we denote these quantities as $\dot\Mwj$), and the total mechanical
energy flux ($\ldwin$ and $\lj$, eqs.~[\ref{eq:ldwin}] and
~[\ref{eq:ldjet}]; denoted as $L_{\rm wj}$). The solid angle of each
cone, that we call in general $\DOmec$ (constant for the jet, and
given by eq.~[\ref{eq:domeg}] for the wind), has a {\it linear}
opening angle $\Thc$, so that $\DOmec=\pi\Thc^2$. From elementary
geometrical considerations, one obtains for the linear opening of the
cone the relation $\Rc (r)= r\Thc(r)$, $\dot\Mwj=\vwj(r)\pi
\Rc^2(r)\rho_{\rm wj}(r)$, and $L_{\rm wj}=0.5\dot\Mwj\vwj^2(r)$.  As
described in Sect.~2.3, the jets were taken to be very narrow and
nearly relativistic, with energy efficiency increasing as $\dot\mbh$
decreases. The winds were taken to have much larger opening angles,
much lower (but still supervirial) velocities and to have an energy
efficiency that increased with increasing $\mdot(t)$ (see Fig.~1 for
details).

Certainly the simplest description that one could propose would be to
imagine that no energy or mass was transferred from the out-flowing
supersonic flow to the ambient fluid until the internal momentum flux
per unit area (declining as $r^{-2}$) reached the level where it
locally matched the pressure in the ambient fluid, and define that
radius ($\Rwj$) to be that of the “working surface” where the
wind or jet dumped all of its mass, momentum and energy. In this
simplest of all models there would be no input from the wind or jet
until this point was reached as the jet/wind was driven loss-lessly
through the ISM of the galaxy.  So, below the working surface, where
$\beta (r)\equiv\Pism/\Pwj <1$, the flow is conservative, but
then as the pressure ratio approaches unity the mass and energy are
deposited rapidly into a thin shell.  While this approach has the
virtue of simplicity and conservation of the vital quantities (and it
was the first that we explored numerically), it is numerically
unattractive and unstable.

Next one could adopt the similar but smoother scheme described in
eq.~(\ref{eq:mechfed}) but without the time dependent term. In this
case there are some losses below and above the working surface, but
the bulk of the mass and energy are dumped into the ambient medium
over approximately one scale-length in radius in the vicinity of
$\Rwj$. This is the equation that we adopted in this paper for
adding energy and mass from the wind/jet to the ambient flow.  This
approach has assumed implicity that all wind/jet outflow velocities
are highly supersonic so that the quantities in the outflow at
distance $r$ from the center and at time $t$ are essentially the same
as at $R_1$ and $t$.

But it is straightforward to allow for propagation effects – the
finite time it takes for the wind/jet to flow from the first active
grid point $R_1$ to $r$. Since the total mass, momentum and energy flux
(integrated over the sphere) are conserved beyond $R_1$ aside from
losses calculated explicitly (“sinks”, indicated in the following with
$S$), we can write the more general equation
\beq
{\partial Y\over\partial t} + \vwj{\partial Y\over\partial r}=-S(r,t).
\label{eq:A1}
\eeq
We then finally choose to write the sink term as $S(r,t)=Y/\tau(r,t)$,
so that $dt/\tau(r,t)$ is the fraction of the outflow mass (or energy
or momentum) flux that is deposited in time $dt$ at
$(r,t)$. Consistent with the concept of depositing the mass, momentum
and energy smoothly in the vicinity of the working surface, we take
\beq
{1\over\tau}={\vwj\over r}{\Pism\over\Pwj}=
{\vwj\over r}\beta(r,t),
\label{eq:A2}
\eeq
where $\beta$ is the pressure ratio.  Inserting eq.~(\ref{eq:A2}) in
eq.~(\ref{eq:A1}), and after some rearrangement, we obtain
eq.~(\ref{eq:mechfed}). It reduces to our first, most simplified
proposal (no explicit time dependent term) in the limits that either
the flow is steady or it is of very large velocity (i.e., $\vwj\gg
{\rm max}(v,\cs)$). The full equation (\ref{eq:mechfed}), which allows
for the lags in a time-dependent flow, provides more stable solutions
but is more costly to implement than the non time-dependent form
because the Courant condition for the time-step must be modified to be
\beq
\Delta t < {\rm min}\left ({\Delta r\over \sqrt{\cs^2 + v^2}},
{\Delta r\over \vwj}\right),
\label{eq:A3}
\eeq
where $\Delta r$ is the spatial grid spacing (for simplicity the
contribution of artificial viscosity is not reported in equation
above).  In this paper we used the more economical time-independent
version rather than the more accurate but costly to implement full
version, which is used and discussed in Paper III.

Now let us back up and ask if, for a steady conical flow, the time
independent equation can be derived by any more formal arguments drawn
from first principles. For a steady conical flow, which interacts with
the ambient flow only at its boundary, it is easy to derive the
following equations from the conservation laws. If the linear opening
angle of the cone is $\Thc$, then from geometrical considerations
\beq
{d\ln\Thc\over d\ln r}= {\cs\over\Thc \vwj}-1,
\label{eq:A4}
\eeq
where the velocity of each fluid element in the outflow perpendicular
to the radial direction is assumed to be the speed of sound $\cs$ in
that flow at radius $r$. The equation of mass conservation is
\beq
{d\ln\rho\over d\ln r}=-2 -{d\ln \vwj\over d\ln r}-2{d\ln\Thc\over d\ln r},
\label{eq:A5}
\eeq
and we are now allowing the flow velocity to be a function of
radius. The First Law of Thermodynamics ($dU=dQ - PdV$) can be written
as
\beq
2{d\ln\cs\over d\ln r}=(\gamma-1){d\ln\rho\over d\ln r} + 
                        {\gamma-1\over\cs^2}{d q\over d\ln r},
\label{eq:A6}
\eeq
where $q(r,t)$ is the heat energy added per unit mass of the flow. The
conservation of energy (neglecting gravitational energy in this
super-virial flow) can be written as 
\beq
{d\ln \vwj\over d\ln r}=-{2\over\gamma-1}{\cs^2\over \vwj^2}+
                           \beta(r){d\ln\rho\over d\ln r},
\label{eq:A7}
\eeq
where we have allowed for the work done on the external 
fluid, as described beelow. We have not required that the pressure in the
wind/jet balances the external pressure and in such a highly
supersonic flow this would not naturally be the case (the jet fluid
tending to rapidly cool due to the adiabatic expansion,
i.e. $\cs^2(r)\rho(r) \ll\Pism$). We can define the {\it equilibrium}
internal speed sound, $\ceq$, that would correspond to pressure
equilibrium across the cone boundary as
\beq
\rho(r)\ceq^2(r)\equiv\Pism(r),
\label{eq:A8}
\eeq
and in general we expect that, if $\cs (r)\ll\ceq (r)$, then the
wind/jet will be unstable to rippling, internal shocks will develop,
and we can plausibly assume that the heat generated by such processes
will result in energy being taken from the bulk flow (reducing $\vwj$)
and being added to the thermal energy in the wind/jet, with the heat
generated by the instability proportional to $\ceq^2(r) -\cs^2(r)$.
Of course, additional phenomena such as strong shear and large density
contrast at the wind/jet boundary are likely to give rise to
instabilities there, regardless of the value of $\cs$. Then, from
dimensional analysis we would obtain
\beq
{dq\over d\ln r}={\rm constant}\times {\ceq\over\vwj}\times
                 [\ceq^2(r)-\cs^2(r)].
\label{eq:A9}
\eeq
Note that the instabilities converting forward
kinetic energy in the wind/jet to thermal energy, are also responsible for
the source terms of mass, momentum and energy transferred from the wind/jet 
to the ISM along the boundaries of the cone.  
The set of six equations (A4)-(A9) would allow one to compute the
radial dependence of the six variables $\rho$, $\vwj$, $\Thc$,
$\ceq$, $\cs$, $q$ (or in principle their more complicated
time-dependent versions) to determine the rate at which energy and
mass are drained from the sides of the wind/jet and from its working
surface. If however the instabilities are very efficient, we can short
circuit this process, drop the $q(r)$ variable and simply assume that
heat is supplied to the outflow (and taken from the bulk motion) at the
rate required to maintain $\cs (r)=\ceq (r)$. Then, omitting several
steps of the straightforward algebra, and approximating the flow by
power-law solutions (i.e., all variables are taken to scale as powers
of the radius $r$), one obtains, in the case of a steady wind/jet
\beq
{d\ln Y\over d\ln r}=-\beta(r)\left [{2\gamma + \mu\over 
                 \gamma -1 +\gamma\beta (r)}\right ],\quad 
                 \mu\equiv {d\ln\Pism\over d\ln r},
\label{eq:A10}
\eeq
which, aside from the bracket is the same as the time-independent
version of equation (\ref{eq:mechfed}) that we use in this paper.
Typical values for the quantities in the brackets, considerably below
the working surface are $\gamma = 5/3$, $\mu = -2$, $\beta\to 0$, so
that the term in square brackets tends to the value of 2, giving a
result which is very close to our time-independent version of
eq.~(29). The essential conclusion of this section is that, while a
full 2D or 3D, time-dependent treatment of the energy and mss
deposition from an out-flowing wind/jet is highly desirable and in
fact essential for a careful, quantitative treatment of this problem,
nevertheless, the simple formulation presented in eq.~(29) does offer
a conservative scheme which captures several of the basic features of
the problem.

\end{appendix}



\begin{thebibliography}

\bibitem[]{} Antonuccio-Delogu, V. \&  Silk, J.  2008,
             \mnras, 389, 1750

\bibitem[]{} Arav, N. et al. 2007, 
             \apj, 658, 829

\bibitem[]{} Begelman, M.C., \& Nath, B.B. 2005,
             \mnras, 361, 1387

\bibitem[]{} Begelman, M.C., \& Ruszkowski, M. 2005,
             Phil.Trans. of Roy.Soc., part A, 363, n.1828, 655

\bibitem[]{} Bertin, G., et al. 1994,
             A\&A, 292, 381

\bibitem[]{} Binney, J. 2001,
             in "Particles and Fields in Radio Galaxies Conference", 
             ASP Conference Proceedings, Robert A. Laing and 
             Katherine M. Blundell eds., vol. 250, p. 481

\bibitem[]{} Binney, J., \& Evans, N.W. 2001,
             \mnras, 327, L27

\bibitem[]{} Binney, J., \& Tabor, G. 1995,
             \mnras, 276, 663

\bibitem[]{} Blustin, A.J., Kriss, G.A., Holczer, T., Behar, E., 
             Kaastra, J.S., Page, M.J., Kaspi, S., Branduardi-Raymont, G., \&
             Steenbrugge, K.C. 2007, 
             A\&A, 466, 107

\bibitem[]{} Burkert, A., \& Silk, J. 2001,
             \apj, 554, L151

\bibitem[]{} Cappellari, M., et al. 2006,
             \mnras, 366, 1126

\bibitem[]{} Castor, J.I., Abbott, D.C., \&  Klein, R.I. 1975,  
             \apj, 195, 157

\bibitem[]{} Cavaliere, A., \& Vittorini, V. 2002,
             \apj, 570, 114

\bibitem[]{} Chartas, G., Brandt, W.N., \& Gallagher, S.C. 2003, 
             \apj, 595, 85

\bibitem[]{} Churazov, E., Sazonov, S., Sunyaev, R., Forman, W., Jones, C., \&
             B\"ohringer, H. 2005,
             \mnras, 363, L91

\bibitem[]{} Ciotti, L. 2009,
             La Rivista del Nuovo Cimento, 32, n.1., 1

\bibitem[]{} Ciotti, L., \& Ostriker, J.P. 1997,
             \apjl, 487, L105 (CO97)

\bibitem[]{} Ciotti, L., \& Ostriker, J.P. 2001, 
             \apj, 551, 131 (CO01)

\bibitem[]{} Ciotti, L., \& Ostriker, J.P. 2007, 
             \apj, 665, 1038 (CO07)

\bibitem[]{} Ciotti L., Morganti, L., \& de Zeeuw, P.T. 2008, 
             \mnras, 393, 491

\bibitem[]{} Cowie, L.L., Ostriker, J.P., \& Stark, A.A. 1978,
             \apj, 226, 1041

\bibitem[]{} Crenshaw, D.M., Kraemer, S.B., \& George, I.M. 2003,
             ARAA, 41, 117

\bibitem[]{} Croton, D.J., et al. 2006,
             \mnras, 365, 11

\bibitem[]{} Czoske, O., Barnabe', M., Koopmans, L.E.V., Treu, T., Bolton, A.S.
             2008, \apj, 384, 987

\bibitem[]{} de Zeeuw, P.T. 2001, in the Proceedings of the ESO workshop 1999
             L. Kaper, E.P.J. van den Heuvel and P.A. Woudt Eds., pag. 78 

\bibitem[]{} Diehl, S., \& Statler, T.S. 2008
             \apj, 687, 986

\bibitem[]{} Di Matteo, T., Springel, V., \& Hernquist, L. 2005,
             Nature, 433, 604

\bibitem[]{} Dorodnitsyn A., Kallman T., \& Proga, D. 2008,
             \apjl, 675, 5

\bibitem[]{} Douglas, N.G., et al. 2007,
             \apj, 664, 257

\bibitem[]{} Dubinski, J., \& Carlberg, R.G. 1991,
             \apj, 378, 496

\bibitem[]{} Dye, S., Evans, N.W., Belokurow, V., Warren, S.J., \& Hewett, P.
             arXiv:0804.4002

\bibitem[]{} Faber, S.M., et al. 1997,
             \aj, 114, 1771

\bibitem[]{} Fabian, A.C. 1999,
             \mnras, 308, L39

\bibitem[]{} Fabian, A.C., Thomas, P.A., Fall, S.M., \& White III, R.E. 1986,
             \mnras, 221, 1049

\bibitem[]{} Ferrarese, L., \& Ford, H. 2005,
             Space Science Reviews, 116, 523

\bibitem[]{} Ferrarese, L., \& Merritt, D. 2000,
             \apj, 539, L9

\bibitem[]{} Fukushige, T., \& Makino, J. 1997,
             \apj, 477, L9

\bibitem[]{} Gallo, E., Treu, T., Jacob, J., Woo, J.H., Marshall, P., \& 
             Antonucci, R. 2008,
             \apj, 680, 154

\bibitem[]{} Ganguly R., \& Brotherton, M.S. 2008,
             \apj, 672, 102

\bibitem[]{} Gavazzi, R., Treu, T., Rhodes, J.D.. Koopmans, L.V.E., Bolton, 
             A.S., Burles, S., Massey, R.J., \& Moustakas, L.A. 2007, 
             \apj, 667, 176

\bibitem[]{} Gebhardt, K., et al. 2000,
             \apj, 539, L13

\bibitem[]{} Ghisellini, G.
             arXiv:0807.2640

\bibitem[]{} Goncalves, T.S., Steidel, C.C., \& Pettini, M. 2008,
             \apj, 676, 816

\bibitem[]{} Graham, A.W., Erwin, P., Caon, N., \& Trujillo, I. 2003,
             Rev.Mex.A.A., 17, 196

\bibitem[]{} Granato, G.L., De Zotti, G., Silva, L., Bressan, A., \& 
             Danese, L. 2004,
             \apj, 600, 580

\bibitem[]{} Haiman, Z., Ciotti, L. \& Ostriker, J.P. 2004,
             \apj, 606, 204

\bibitem[]{} Hamann, F., Kaplan, K.F., Rodriguez Hidalgo, P., Prochaska, J.X., 
             \& Herbert-Fort, S. 2008,
             \mnras, 391, L39

\bibitem[]{} Helmboldt, J.F., Walterbos, R.A.M., \& Goto, T. 2008,
             \mnras, 387, 1537

\bibitem[]{} Hernquist, L. 1990,
             \apj, 356, 359

\bibitem[]{} Ho, L.C. 2008
             ARAA, 46, 475

\bibitem[]{} Holt, J., Tadhunter, C.N., \& Morganti, R. 2008,
             \mnras, 387, 639

\bibitem[]{} Hopkins, P.F., \& Hernquist, L.
             arXiv:0809.3789

\bibitem[]{} Hopkins, P.F., Hernquist, L., Cox, T.J., Robertson, B., 
             Di Matteo, T., \& Springel, V. 2006,
             \apj, 639, 700

\bibitem[]{} Hopkins, P.F., Cox, T.J., Dutta, S.N., Hernquist, L, 
             Kormendy, J, \& Lauer, T.R. 2008a
             arXiv:0805.3533

\bibitem[]{} Hopkins, P.F., Lauer, T.R., Cox, T.J., Hernquist, L., \&
             Kormendy, J. 2008b
             arXiv:0806.2325

\bibitem[]{} Humphrey, P.J., Buote, D.A., Gastaldello, F., Zappacosta, L., 
             Bullock, J.S., Brighenti, F., \& Mathews, W.G. 2006,
             \apj, 646, 899

\bibitem[]{} Jaffe, W. 1983
             \mnras, 202, 995

\bibitem[]{} Jaffe, W., Ford, H.C., O'Connell, R.W., van den Bosch, F.C., \&
             Ferrarese, L. 1994, 
             \aj, 108, 1567

\bibitem[]{} Johansson, P.H., Naab, T., \& Burkert, A. 2008,
             Astron.Nachr. 329, 956

\bibitem[]{} Jolley, E, \&  Kuncic, Z.
             arXiv:0802.0902

\bibitem[]{} Kauffman, G., \& Heckman, T.M.
             arXiv:0812.1224

\bibitem[]{} King, I.R. 1972, 
             \apj, 174, L123

\bibitem[]{} King, A.R. 2003, 
             \apj, 596, L27

\bibitem[]{} Kirkman, D., \& Tytler, D. 2008,
             \mnras, 391, 1457

\bibitem[]{} Koopmans, L.V.E., Treu, T., Bolton, A.S.,
             Burles, S., \& Moustakas, L.A. 2006, 
             \apj, 649, 599

\bibitem[]{} K\"onigl A. 2006, 
             MemSAIt, 77, 598

\bibitem[]{} Kormendy, J., \& Richstone, D. 1995,
             ARAA, 33, 581

\bibitem[]{} Kurosawa, R. \& Proga, D. 2008a,
             \apj, 674, 97

\bibitem[]{} Kurosawa,  \& Proga, D. 2008b,
             arXiv:0812.3153

\bibitem[]{} Lauer, T.R., et al. 2005,
             \aj, 129, 2138

\bibitem[]{} Li, C., Kauffmann, G., Heckman, T.M., White, S.D.M., \& 
             Jing, Y.P. 2008,
             \mnras, 385 1915

\bibitem[]{} Magorrian, J., et al. 1998,
             \aj, 115, 2285

\bibitem[]{} Marconi, A. \& Hunt, L.K. 2003,
             \apj, 589, L21

\bibitem[]{} Martinez-Sansigre, A., \& Taylor, A.M.
             arXiv:0810.3920

\bibitem[]{} Martini, P. 2004, in
             Coevolution of Black Holes and Galaxies, ed. L.C. Ho, 169

\bibitem[]{} McCarthy, I.G., Babul, A., Bower, R.G., \& Balogh, M.L. 2008,
             \mnras, 386, 1309

\bibitem[]{} McLure, R.J., \& Dunlop, J.S. 2002,
             \mnras, 331, 795

\bibitem[]{} McNamara, B.R, \& Nulsen, P.E.J. 2007,
             ARAA, 45, 117

\bibitem[]{} Merloni, A. \&  Heinz, S. 2008
             \mnras, 388, 1011

\bibitem[]{} Milosavljevic, M., Bromm, V., Couch, S.M., \& Oh, S.P.
             arXiv0809.2404

\bibitem[]{} Morgan, C.W., Kochanek, C.S., Morgan, N.D., \& Falco, E.M.
             arXiv0707.0305

\bibitem[]{} Murray, N., Quataert, E., \& Thompson, T.A. 2005,
             \apj, 618, 569

\bibitem[]{} Narayan, R. \& Yi, I. 1994
             \apj, 428, L13

\bibitem[]{} Navarro, J.F., Frenk, C.S., \& White, S.D.M. 1997,
             \apj, 490, 493

\bibitem[]{} Nayakshin, S., \& Sunyaev, R. 2005,
             \mnras, 364, L23

\bibitem[]{} Nayakshin, S., Dehnen, W., Cuadra, J., \& Genzel, R. 2006,
             \mnras, 366, 1410

\bibitem[]{} Noble, S.C., Krolik, J.H., \& Hawley, J.F.
             arXiv:0808.3140

\bibitem[]{} Omma, H., Binney, J., Bryan, G., \& Slyz, A. 2004, 
             \mnras, 348, 1105

\bibitem[]{} O'Sullivan, E., Ponman, T.J., \& Collins, R.S. 2003,
             \mnras, 340, 1375   

\bibitem[]{} Ostriker, J.P., \& Ciotti, L. 2005,
             Phil.Trans. of Roy.Soc., part A, 363, n.1828, 667 (OC05)

\bibitem[]{} Parriott, J.R., \&  Bregman, J.N. 2008,
             \apj, 681, 1215

\bibitem[]{} Pellegrini, S. 2005a,
             \apj, 624, 155

\bibitem[]{} Pellegrini, S. 2005b,
             \mnras, 364, 169

\bibitem[]{} Pellegrini, S., \& Ciotti, L. 1998,
             A\&A, 333, 433

\bibitem[]{} Pellegrini, S., Ciotti, L. \& Ostriker, J.P. 2009,
             Advances in Space Research, in press

\bibitem[]{} Peterson, J.R., \& Fabian, A.C. 2006,
             Phys.Rep., 427, 1

\bibitem[]{} Pierce, C.M., et al. 2007,
             \apj, 660, L19

\bibitem[]{} Pipino, A., Silk, J., \&  Matteucci, F.
             arXiv:0810.2045

\bibitem[]{} Pipino, A., Kaviraj, S., Bildfell, C., Hoekstra, H.,  Babul, A., 
             \& Silk, J. 
             arXiv:0807.2760

\bibitem[]{} Prochaska, J.X., \& Hennawi, J.F.
             arXiv:0806.0862

\bibitem[]{} Proga, D. 1999, 
             \mnras, 304, 938

\bibitem[]{} Proga, D. 2007a,
             \apj, 661, 693

\bibitem[]{} Proga, D. 2007b, in The Central Engine of Active Galactic Nuclei,
             eds. L.C. Ho, \& J.M. Wang, ASP Conf. Ser., 373, 267

\bibitem[]{} Proga, D., \&  Kallman T. 2002,
             \apj, 565, 455

\bibitem[]{} Proga, D., \&  Kallman T. 2004, 
             \apj, 616, 688

\bibitem[]{} Proga, D., Stone J.M., \& Drew J.E. 1998, 
             \mnras, 295, 595

\bibitem[]{} Proga, D., Stone J.M., \& Kallman T. 2000, 
             \apj, 543, 686

\bibitem[]{} Proga, D., Ostriker, J.P., \& Kurosawa, R. 2008, 
             \apj, 676, 101

\bibitem[]{} Rafferty, D.A., McNamara, B.R., \& Nulsen, P.E.J. 2008,
             \apj, 687, 899

\bibitem[]{} Renzini, A., Ciotti, L., D'Ercole, A. \& Pellegrini, S. 1993,
             \apj, 419, 52

\bibitem[]{} Riciputi, A. Lanzoni, B., Bonoli, S., \&  Ciotti, L. 2005,
             A\&A, 443, 133

\bibitem[]{} Rusin, D. \& Kochanek, C.S. 2005, 
             \apj, 623, 666

\bibitem[]{} Rusin, D., Kochanek, C.S., \& Keeton, C.R. 2003, 
             \apj, 595, 29

\bibitem[]{} Saglia, R.P., et al. 1993, 
             A\&A, 403, 567

\bibitem[]{} Sazonov, S.Yu., Ostriker, J.P., \& Sunyaev, R. 2004,
             \mnras, 347, 144

\bibitem[]{} Sazonov, S.Yu., Ostriker, J.P., Ciotti, L., \& Sunyaev, R.A. 
             2005, \mnras, 358, 168

\bibitem[]{} Sazonov, S.Yu., Revnivtsev, M., Krivonos, R., Churazov, E., \& 
             Sunyaev, R.A. 2007, 
             A\&A, 462, 57

\bibitem[]{} Sazonov, S.Yu., Krivonos, R., Revnivtsev, M., Churazov, E., \& 
             Sunyaev, R.A. 2008, 
             A\&A, 482, 517

\bibitem[]{} Schawinski, K., Lintott, C.J.,  Thomas, D., Kaviraj, S., 
             Viti, S.,  Silk, J., Maraston, C., Sarzi, M.,  Yi, S.K.,  
             Joo, S.J., Daddi, E., Bayet, E.,  Bell, T., \& Zuntz, J. 
             arXiv:0809.1096 

\bibitem[]{} Shakura N.I., Sunyaev R.A. 1973, 
             A\&A, 24, 337

\bibitem[]{} Shi J. \& Krolik J. 2008,
             \apj, 679, 1018

\bibitem[]{} Sijacki, D., Springel, V., Di Matteo, T., \& Hernquist, L. 2007, 
             \mnras, 380, 877

\bibitem[]{} Sijacki, D., Pfrommer, C., Springel, V., \&  Ensslin, T.A. 2008,
             \mnras, 387, 1403

\bibitem[]{} Silk, J., \& Rees, M.J. 1998,
             A\&A, 331, L1

\bibitem[]{} Soltan, A. 1982,
             \mnras, 200, 115

\bibitem[]{} Somerville, R.S. 2008,
             in Panoramic views of galaxy formation and evolution, 
             ASP Conference Series, vol.399, eds. T. Kodama, T. Yamada, 
             K. Aoki, p. 391

\bibitem[]{} Springel, V., Di Matteo, T. \& Hernquist, L. 2005,
             \mnras, 361, 776

\bibitem[]{} Tabor, G.,\& Binney, J. 1993
             \mnras, 263, 323

\bibitem[]{} Trager, S.C., Faber, S.M., Dressler, A. 2008,
             \mnras, 386, 715

\bibitem[]{} Treu, T. \& Koopmans, L.V.E. 2002, 
             \apj, 575, 87 

\bibitem[]{} Treu, T. \& Koopmans, L.V.E. 2004, 
             \apj, 611, 739

\bibitem[]{} Turnshek, D.A. 1988, in QSO Absorption Lines: 
             Probing the Universe, ed. J.C. Blades, D.A. Turnshek, \& 
             C.A. Norman (Cambridge: Cambridge Univ. Press), 17

\bibitem[]{} Vernaleo, J.C. \& Reynolds, C.S. 2006,
             \apj, 645, 83

\bibitem[]{} Voit, G.M., \& Donahue, M. 2005,
             \apj, 634, 955

\bibitem[]{} Wang, J.M. 2008,
             \apj, 682, L81

\bibitem[]{} Watabe Y., Kawakatu, N., \& Imanishi M. 2008,
             \apj, 677, 895

\bibitem[]{} Woo, J.H., Treu, T., Malkan, M.A., \& Blandford, R.D.
             arXiv:0804.0235

\bibitem[]{} Wrobel, J.M.,  Terashima, Y., \& Ho, L.C. 2008,
             \apj, 675, 1041

\bibitem[]{} Wyithe, J.S.B., \& Loeb, A. 2003,
             \apj, 595, 614

\bibitem[]{} Yu, Q., \& Tremaine, S. 2002, 
             \mnras, 335, 965

\end{thebibliography}
\end{document}